\documentclass[
aps,nofootinbib,showpacs,showkeys,preprint
,tightenlines ] {revtex4}

\usepackage{epsf,epsfig,subfigure,graphicx,amsmath,amssymb}
\usepackage{color}

\newcommand{\dis}[1]{\begin{equation}\begin{split}#1\end{split}\end{equation}}
\newcommand{\be}{\begin{equation}}
\newcommand{\ee}{\end{equation}}

\newcommand{\eq}[1]{Eq.~(\ref{#1})}

\newcommand{\VEV}[1]{\langle #1 \rangle}

\def\bea{\begin{eqnarray}}
\def\eea{\end{eqnarray}}

\begin{document}

\title{\Large\bf 
Inert Higgs extension of the NMSSM 
}

\author{Bumseok Kyae\footnote{email: bkyae@pusan.ac.kr}
}
\affiliation{
Department of Physics, Pusan National University, Busan 609-735, Korea
}


\begin{abstract}

We introduce one pair of inert Higgs doublets $\{H_d,H_u\}$ and singlets $\{N^c,N\}$, and  
consider their couplings with the Higgs doublets of the minimal supersymmetric standard model (MSSM), 
$W\supset y_NN^ch_uH_d+y_N^\prime Nh_dH_u$. 
We assign extra U(1)$_{Z^\prime}$ gauge charges only to the extra vector-like superfields, 
and so all the MSSM superfields remain neutral under the new  U(1)$_{Z^\prime}$. 
They can be an extension of the ``$\lambda$ term,'' $W\supset\lambda Sh_uh_d$ in the next-to-MSSM (NMSSM). Due to the U(1)$_{Z^\prime}$, the maximally allowed low energy value of $y_N$ can be lifted up to $0.85$, avoiding a Landau-pole (LP) below the grand unification scale. 
Such colorless vector-like superfields remarkably enhance the radiative MSSM Higgs mass particularly for large ${\rm tan}\beta$ through the $y_N$ term and the corresponding holomorphic soft term. 
As a result, the lower bound of $\lambda$ and the upper bound of ${\rm tan}\beta$ can be relaxed to disappear  
from the restricted parameter space of the original NMSSM, $0.6\lesssim\lambda\lesssim 0.7$ and $1<{\rm tan}\beta\lesssim 3$. 
Thus, the valid parameter space significantly expands up to $0\lesssim\lambda\lesssim 0.7$, $0\lesssim y_N\lesssim 0.85$, and $2\lesssim {\rm tan}\beta\lesssim 50$, 
evading the LP problem and also explaining the 126 GeV Higgs mass naturally.   

\end{abstract}


\maketitle

\section{Introduction}

%

Recently, ATLAS and CMS collaborations have discovered the Higgs boson with a mass around 126 GeV \cite{LHC}, which would be a triumph of the standard model (SM). 
Thus, the status of the SM as the basic theory describing the nature becomes further stabilized.  
Since the models based on low energy supersymmetry (SUSY) predicted a relatively light Higgs mass, one might say that the 126 GeV Higgs boson supports also SUSY.  
Unfortunately, however, any evidence of new physics beyond the SM including SUSY 
has not been  observed yet at the large hadron collider (LHC). It implies that the theoretical puzzles raised in the SM such as the gauge hierarchy problem, which have provided motivations of new physics 
for last four decades, still remain unsettled.

In fact, 126 GeV is too heavy for a mass of the Higgs appearing in the minimal supersymmetric SM (MSSM).   
The basic reason of it is that the tree-level quartic coupling of the MSSM Higgs potential is given by the small SM gauge couplings unlike the SM.  
As a consequence, the predicted tree-level Higgs mass in the MSSM is lighter even than the $Z$ boson mass $M_Z$ \cite{book,twoloop}. 
Thus, a large radiative correction for lifting the Higgs mass is very essential in the MSSM to account for the observed Higgs mass.

The dominant radiative correction to the Higgs mass  in the MSSM comes from the top quark Yukawa coupling  in the superpotential ($W\supset y_tQ_3h_uu_3^c$), 
because the relevant Yukawa coupling constant is large enough ($y_t\approx 1$). 
Through the top quark Yukawa coupling, 
the top quark and its super-partner, ``stop'' contribute to the radiative Higgs mass ($\equiv\Delta m_h^2|_{\rm top}$) and also the renormalization of the soft mass parameter of the Higgs $h_u$ ($\equiv\Delta \widetilde{m}_2^2|_{\rm top}$) \cite{book,twoloop}:
\dis{ \label{TopStop}
&\Delta m_h^2|_{\rm top}\approx 
\frac{3m_t^4}{4\pi^2v_h^2}\left[{\rm log}\left(\frac{\widetilde{m}_t^2}{m_t^2}\right) 
+\frac{A_t^2}{\widetilde{m}_t^2}\left(1-\frac{1}{12}\frac{A_t^2}{\widetilde{m}_t^2}\right)\right] , 
\\
&  \quad
\Delta \widetilde{m}_2^2|_{\rm top}\approx \frac{3|y_t|^2}{8\pi^2}\widetilde{m}_t^2 
~{\rm log}\left(
\frac{\widetilde{m}_t^2}{M_G^2}\right)
\left[1 + \frac12\frac{A_t^2}{\widetilde{m}_t^2}\right] ,
}
where $m_t$ ($\widetilde{m}_t$) denotes the top quark (stop) mass, and $v_h$ 
is the vacuum expectation value (VEV) of the Higgs, $v_h\equiv\sqrt{\VEV{h_u}^2+\VEV{h_d}^2}\approx 174~{\rm GeV}$ with $\tan\beta\equiv\VEV{h_u}/\VEV{h_d}$. 
$M_G$ means the grand unified theory (GUT) scale ($\approx 2\times 10^{16}~{\rm GeV}$) adopted as a cut-off of the model.     
For simple expressions, here we assumed that the ``$A$-term'' coefficient corresponding to the top quark Yukawa coupling,  
$A_t$ dominates over $\mu \cdot{\rm cot}\beta$, 
where $\mu$ stands for the ``$\mu$-term'' coefficient in the MSSM superpotential.\footnote{The simple expressions of \eq{TopStop} are obtained, 
when the SU(2)$_L$-doublet and -singlet stops are degenerate. 
However, they would be good approximations, 
unless the stops are too hierarchical to be realized in supergravity (SUGRA) models. 
For the full expressions, refer to e.g. Ref.~\cite{book}.}
As seen in \eq{TopStop}, only useful parameters for enhancing the radiative Higgs mass, $\Delta m_h^2|_{\rm top}$ are $\widetilde{m}_t^2$ and $A_t^2$ in the MSSM. 
For the 126 GeV Higgs mass, it is known that a stop mass should be heavier than at least a few TeV with the help of the $A_t^2$ term, 
if two-loop effects are also included \cite{twoloop}.  
Note that the radiative Higgs mass squared, $\Delta m_h^2|_{\rm top}$ is maximized 
when $A_t^2$ fulfills $A_t^2/\widetilde{m}_t^2=6$ \cite{book,twoloop}.  

The renormalization contribution in \eq{TopStop}, $\Delta \widetilde{m}_2^2|_{\rm top}$ is associated with the fine-tuning problem in the Higgs sector, 
since the Higgs VEV  
and the resulting $Z$ boson mass,  
which define the electroweak (EW) scale, are determined by $\widetilde{m}_2^2$ ($\approx\widetilde{m}_2^2|_{\rm tree}+\Delta \widetilde{m}_2^2|_{\rm top}$)
[and other (soft) mass parameters].  
If $\widetilde{m}_t^2$ is of order TeV for the 126 GeV Higgs mass, thus, a TeV scale fine-tuning among the soft parameters is needed to obtain the $Z$ boson mass of $91~{\rm GeV}$ (``little hierarchy problem'').     
Particularly, the ``maximal mixing'' case, $A_t^2/\widetilde{m}_t^2=6$ makes $\Delta \widetilde{m}_h^2|_{\rm top}$ four times larger than the case of $A_t^2\approx 0$, which aggravates the fine-tuning problem.     

In order to mitigate the fine-tuning in the Higgs sector, thus, the stop mass squared and $A_t$ should be as small as possible. 
At the moment, the stop mass bound is $\gtrsim 700~{\rm GeV}$ \cite{stopmass}.\footnote{For the lightest super-particle (LSP) heavier than $250-300~{\rm GeV}$, the stop mass is not constrained at all. 
If $R$ parity is broken, the lower bounds on stop and gluino masses can be relieved to $\gtrsim 500-600~{\rm GeV}$ \cite{Rviol}.}  
However, the radiative correction by a heavy gluino ($\gtrsim 1.4 ~{\rm TeV}$ \cite{gluinomass}) pulls up $\widetilde{m}_t^2$ too much at the EW scale. 
Thus, the heavy gluino effect should be adequately  suppressed 
for a light stop mass at low energy. 
Actually it could be compensated e.g. by very heavy squarks of the first and second generations ($\sim 10~{\rm TeV}$) \cite{naturalSUSYU(1)medi}\footnote{This mechanism would demand another fine-tuning \cite{Hardy,Rviol}, unless an elaborate model is contrived. However, we will not pursue that ambitious goal in this paper.}
through two loop effects \cite{naturalSUSY}, 
or by the Wino but giving up the assumption of the universal gaugino mass at the GUT scale.\footnote{This idea could be applied to achieve a small enough $\widetilde{m}_2^2$ at the EW scale \cite{non-universalgaugino}.}
Otherwise, one should assume a low messenger scale 
to minimize the renormalization group (RG) effect of a heavy gluino.  
In this paper, we don't specify a scenario for a small enough stop mass. 
Although one somehow acquires a small stop mass and successfully reduces the fine-tuning, however, 
one cannot explain yet the observed Higgs mass just with a small stop mass. 
With $\widetilde{m}_t^2\gtrsim (700~{\rm GeV})^2\gg A_t^2$, we have just $\Delta m_h^2\gtrsim (79~{\rm TeV})^2$. 
For the 126 GeV Higgs mass, hence, the mass gap, $\Delta m_h^2|_{\rm new}=(36-81~{\rm GeV})^2$ for $50>{\rm tan}\beta >2$ should somehow be filled up. 
For raising the Higgs mass, thus, we need 
other ingredients rather than the stop mass.

As pointed out above, a too light tree-level Higgs mass could be a crucial cause of the Higgs mass problem in the MSSM.  
In the next-to MSSM (NMSSM), the ``$\mu$ term'' of the MSSM ($W\supset \mu h_uh_d$) is promoted to the renormalizable coupling with a new singlet superfield $S$ \cite{NMSSMreview,singletEXT}: 
\dis{ \label{NMSSMsuperPot}
W\supset \lambda Sh_uh_d . 
}
%
%
It provides an additional quartic term in the Higgs potential,  
$V_h\supset |\lambda h_uh_d|^2$, which is absent in the MSSM, 
and so the Higgs mass in the NMSSM receives a {\it tree-level} correction resulting from that term:  
\dis{ \label{NMSSMhiggs}
m_h^2|_{\rm NMSSM} \supset \lambda^2 v_h^2{\rm sin}^22\beta . 
}
Accordingly, the Higgs mass could be significantly increased even with a relatively light stop, if the dimensionless coupling $\lambda$ was of order unity. 
For $\widetilde{m}_t^2\lesssim (700~{\rm GeV})^2$, $\lambda$ should be larger than $0.6$ 
with $1\lesssim {\rm tan}\beta\lesssim 3$ 
for explaining the observed Higgs mass \cite{nmssm2}.
However, $\lambda$ greater than $0.7$ at the EW scale turns out to bring a Landau-pole (LP) below the GUT scale through its RG evolution \cite{Masip}.
%
Thus, only the quite narrow bands of the parameter space are left in the NMSSM: 
\dis{ \label{NMSSMpara}
0.6\lesssim \lambda\lesssim 0.7 ,~~{\rm and}~~ 1< {\rm tan}\beta\lesssim 3 .
}
Note that $\lambda\approx 0.7$ is the upper bound for  perturbativity of the model up to the GUT scale, 
and ${\rm tan}\beta\approx 1$ maximizes \eq{NMSSMhiggs}.  
They imply that the NMSSM accounts for the Higgs mass 
just {\it around the boundary of the theoretically valid parameter space}. 
For naturalness of the Higgs mass in the NMSSM, therefore, the permitted parameter space should be somehow enlarged. 

One way to relieve the LP constraint on $\lambda$ 
is to introduce a new gauge symmetry, under which $S$ and $\{h_u,h_d\}$ are charged:   
a strong enough new gauge interaction could hold  $\lambda$ within the perturbative regime up to the GUT scale, 
and so the upper bound of $\lambda$ at low energy could be relaxed.
In Ref.~\cite{U(1)NMSSM}, a U(1) gauge symmetry is considered for ameliorating the LP problem. 
Since new U(1) gauge charges are assigned to $\{h_u,h_d\}$, however, ordinary MSSM matter fields should also carry proper U(1) charges for the desired Yukawa couplings. 
As a result, the beta function coefficient of the new U(1) gauge coupling becomes much larger, 
which makes the U(1) gauge coupling quite smaller at low energies, and so 
the relaxation mechanism of the LP constraint becomes  inefficient.\footnote{In order to get a small enough beta function coefficient of the U(1) gauge coupling, 
the U(1) charges could be assigned only to one generation, say, the third generation among the MSSM matter fields. For desired Yukawa couplings, in this case an elaborate model should be constructed \cite{U(1)NMSSM}.}     
%

In Ref.~\cite{Vlep}, the Yukawa couplings between the newly introduced vector-like leptons and the ordinary MSSM Higgs were considered for raising the {\it radiative} Higgs mass:
\dis{ \label{Vlep}
W_{\rm VLEP} =  y_{N}Lh_uN^c + y_N^\prime L^ch_dN  ,
+\mu_L LL^c +   \mu_NNN^c  ,
}
where $\{L,L^c\}$ ($\{N^c,N\}$) denote extra lepton doublets (neutral singlets), and 
the last two terms are the mass terms for them \cite{extramatt,extramatt2,JSW}. 
Only if $\{L,L^c\}$ are heavier than $\{N^c,N\}$, 
the former can immediately decay to the latter and SM fermions (plus LSP). 
So they don't leave experimentally unacceptable  signals. 
Since $\{N^c,N\}$ remain absolutely stable, 
they can be a promising dark matter candidate \cite{VlepDM}.  
The Higgs VEVs are aligned in the directions of the neutral components of $\{L,L^c\}$, and so there is no extra two photon enhancement in this model.  
%
Note that the gauge quantum numbers of $\{L,L^c\}$ are the same as those of the MSSM Higgs $\{h_d,h_u\}$, 
and so the structures of the $y_N$ and $y_N^\prime$ terms in \eq{Vlep} are the same as the $\lambda$ term in \eq{NMSSMhiggs}. 
Accordingly, the same LP constraint on $\lambda$ ($\lesssim 0.7$) could be imposed also to $y_N$ and $y_N^\prime$.  

Since one MSSM superfield, $h_u$ ($h_d$) couples to two extra superfields $\{L,N^c\}$ ($\{L^c,N\}$) 
in the $y_N$ ($y_N^\prime$) term of \eq{Vlep}, 
an extra gauge symmetry can be introduced, 
under which {\it all the MSSM superfields remain neutral}, and only the extra vector-like leptons, $\{L,L^c; N^c,N\}$ carry non-trivial extra gauge charges. 
Then, the mixings between the extra vector-like leptons and the ordinary MSSM leptons are forbidden,  
and so such extra lepton doublets don't raise phenomenological problems associated with the flavor changing neutral current (FCNC). 
In Ref.~\cite{Vlep}, the smallest non-Abelian gauge group, SU(2)$_{Z^\prime}$ was considered, which is still {\it asymptotically free}. 
So at least two pairs of vector-like leptons are necessary  
for composing SU(2)$_{Z^\prime}$ doublets. 
In this case, the maximally allowed value of $y_N$ at low energy is significantly lifted up, $|y_N|\lesssim 1.78$.\footnote{For gauge coupling unification, the vector-like leptons should be supplemented with  colored particles, $\{D^c,D\}$. 
In total, three pairs of $\{L,D^c;L^c,D\}$ are   introduced in Ref.~\cite{Vlep}, 
where $\{D^c,D\}$ and one pair of $\{L,L^c\}$ are SU(2)$_{Z^\prime}$ singlets.} 
Only if $|y_N|$ is smaller than $1.78$ at the EW scale, thus, it does not blow up below the GUT scale.

Like the top and stop in the MSSM,   
$\{L,N^c\}$ make contributions to the radiative Higgs mass ($\equiv\Delta m_h^2|_{L,N^c}$) as well as the renormalization of the soft mass squared of $h_u$ ($\equiv\Delta \widetilde{m}_2^2|_{L,N^c}$) \cite{Vlep}:
\dis{ \label{VlepRadi}
&~~ \Delta m_h^2|_{L,N^c}\approx N_V\frac{|y_N|^4}{4\pi^2}v_h^2{\rm sin}^4\beta
~{\rm log}\left(\frac{{M}^2+\widetilde{m}_l^2}{{M}^2}\right) ,
\\
&\Delta \widetilde{m}_2^2|_{L,N^c}\approx N_V\frac{|y_N|^2}{8\pi^2}
\bigg[F_Q({M}^2+\widetilde{m}_l^2)-F_Q({M}^2)\bigg]_{Q=M_G} ,
}
where $N_V=2$ for the ${\rm SU(2)}_{Z^\prime}$ doublets, 
and $F_Q(m^2)$ is defined as $F_Q(m^2)\equiv m^2\{{\rm log}(\frac{m^2}{Q^2})-1\}$. 
$M^2$ denotes the mass squared of the fermionic component, $M^2\approx |\mu_L|^2+|y_N|^2v_h^2\sin^2\beta$.
For simplicity, all the soft mass squareds of $\{L,L^c;N^c,N\}$ are set equal to $\widetilde{m}_l^2$. 
Here $y_N\gg y_N^\prime$ is assumed. 
%
%
Note that $\Delta m_h^2|_{L,N^c}$ is proportional to $|y_N|^4\times v_h^2\sin^4\beta$, 
while $\Delta \widetilde{m}_2^2|_{L,N^c}$
is to $|y_N|^2\times\{M^2+\widetilde{m}_l^2,M^2\}$.
As in the MSSM, $\Delta \widetilde{m}_2^2|_{L,N^c}$ is associated with the fine-tuning of the EW scale. 
For raising the radiative Higgs mass $\Delta m_h^2|_{L,N^c}$, but holding $\Delta \widetilde{m}_2^2|_{L,N^c}$, therefore, a larger $y_N$ and smaller masses of $\{\widetilde{m}_l^2,|\mu_L|^2\}$ are preferred. 
In Ref.~\cite{Vlep}, it was assumed that $|\mu_L|^2$ and the soft parameter $\widetilde{m}_l^2$ are smaller than $\widetilde{m}_t^2$.  
Actually it is possible, because experimental bounds on the leptonic particles are not severe yet.  
The {\it quartic} power of $y_N$ in $\Delta m_h^2|_{L,N^c}$ lifts the radiative Higgs mass very efficiently, if $|y_N|$ is larger than unity. 
Even for the stop mass squared of $\widetilde{m}_t^2\approx (700~{\rm GeV})^2$, thus,     
126 GeV Higgs mass can be easily explained,
only if the following relation is fulfilled: 
\dis{ \label{requirement} 
N_V|y_N|^4 
~{\rm log}\left(\frac{{M}^2+\widetilde{m}_l^2}{{M}^2}\right) ~ \approx ~  13.2,~4.4,~2.8,~2.0,~1.6
}
for ${\rm tan}\beta=2,4,6,10,50$, respectively,\footnote{In Ref.~\cite{Vlep}, the analysis was  carried out with $\widetilde{m}_t^2\approx (500~{\rm GeV})^2$. 
Here we list the numbers estimated with  $\widetilde{m}_t^2\approx (700~{\rm GeV})^2$.} 
even without $A$-term contributions. 
We note that $|y_N|_{\rm max}\approx 1.78$ (or $N_V|y_N|_{\rm max}^4\approx 20$) makes \eq{requirement} trivial. 

%
%

%
%

In this paper, we will introduce an Abelian gauge symmetry U(1)$_{Z^\prime}$ instead of SU(2)$_{Z^\prime}$ for relaxing the LP problem associated with $y_N$.  
Hence, only one pair of vector-like leptons,  $\{L,L^c;N^c,N\}$ would be enough, because they don't have to compose non-trivial multi-plets as in the case of a non-Abelian gauge group. 
Consequently, the model could be much simplified. 
However, since the U(1)$_{Z^\prime}$ gauge coupling, $g_{Z^\prime}$ monotonically increases with energy unlike the non-Abelian case, 
it is hard to get a relatively large $g_{Z^\prime}$ at low energies.  
It means that the relaxation mechanism of the LP problem using an Abelian gauge symmetry would not be much efficient.    
Due to the reason, $y_N$ cannot be large enough to explain 126 GeV Higgs mass with $\widetilde{m}_t^2\gtrsim (700~{\rm GeV})^2$.  
So we will consider also the SUSY breaking $A$-term corresponding to \eq{Vlep} as well as the $\lambda$ 
coupling \eq{NMSSMsuperPot} of the NMSSM.  
Even with a relatively smaller value of $y_N$ ($\lesssim 1$), thus, 
the radiative Higgs mass could be sufficiently raised.  
Most of all, we will attempt to investigate how much the parameter space of $\lambda$ and ${\rm tan}\beta$ can be enlarged in this setup, compared to the case of the original NMSSM. 

In fact, the SM gauge quantum numbers of $\{L,L^c\}$ 
are the same as the MSSM Higgs doublets, $\{h_d,h_u\}$, as mentioned above. 
Thus, introduction of such extra vector-like leptons is equivalent to introduction of new ``inert Higgs doublets'' $\{H_d,H_u\}$, only if the new Higgs doublets don't develop VEVs. 
Hence, the $y_N$ and $y_N^\prime$ terms in \eq{Vlep} can be regarded as extensions of the $\lambda$ term in \eq{NMSSMsuperPot} of the NMSSM 
except for the fact that $\{H_d,H_u\}$ carry extra U(1)$_{Z^\prime}$ charges unlike the ordinary MSSM Higgs $\{h_d,h_u\}$.  
As pointed out above, the U(1)$_{Z^\prime}$ is necessary also for avoiding unwanted FCNC. 
%
In this paper, we will look upon the extra leptons of Ref.~\cite{Vlep} as extra inert Higgs doublets. 

This paper is organized as follows. 
In section \ref{sec:model}, we will discuss the radiative Higgs mass, and the fine-tuning in this model. 
Particularly, we will investigate the $A$-term effects in section \ref{sec:model}.  
In section \ref{sec:LP}, we will analyze the LP constraint on the coupling constants, $y_N$ and $\lambda$, and explore the allowed parameter space. 
We will also compare our results with the cases in the  absence of $y_N$ or $\lambda$ couplings, or the gauged U(1)$_{Z^\prime}$ in section \ref{sec:LP}.    
Section \ref{sec:conclusion} is a conclusion.

\section{Extension of the NMSSM} \label{sec:model}

By introducing one extra pair of the Higgs doublets $\{H_d,H_u\}$ and singlets $\{N^c,N\}$, 
we extend the NMSSM Higgs sector \eq{NMSSMsuperPot} as follows: 
\dis{ \label{superPot}
W_{\rm Hext}=\lambda Sh_uh_d + y_NN^ch_uH_d + y_N^\prime Nh_dH_u +\mu_HH_uH_d + \mu_NNN^c .   
} 
From the last two terms, $\{H_d,H_u\}$ and $\{N^c,N\}$ acquire the masses.  
$\mu_{H}$ and $\mu_N$ of order EW scale can be naturally induced through the Pecci-Quinn symmetry breaking \cite{Kim-Nilles} or SUSY breaking mechanism \cite{GM}.
We assume $\mu_H \gtrsim\mu_N$ 
such that $\{H_d,H_u\}$ immediately decay into $\{N^c,N\}$ plus SM fermions (and LSP) as in \eq{Vlep}. 
In fact, $\mu_N$ is not essential, because $\{N^c,N\}$
can get the masses also from the $y_N$ and $y_N^\prime$ terms, when the Higgs VEVs, $\langle h_u\rangle$ and $\langle h_d\rangle$ are developed. 
$\mu_N$ can be even smaller than $100~{\rm GeV}$.
%
%
Due to the relatively heavy masses of $\{H_d,H_u\}$, 
we assume that $\{H_d,H_u\}$ don't get non-zero VEVs (``inert Higgs doublet''). 
$S$ also get a mass and a VEV by including its self-couplings in the superpotential, and also their soft terms that we don't specify here \cite{NMSSMreview}. 

Actually, the $y_N^\prime$ term ($h_d$ term in general) is less helpful for raising the Higgs mass, since its contribution to the radiative Higgs mass would be ${\rm tan}\beta$ suppressed for ${\rm tan}\beta >1$. 
For a simple analysis, thus, we will neglect $y_N^\prime$ coupling as in \eq{Vlep}, 
assuming $y_N^\prime\ll y_N$.  
Accordingly, in this paper we will consider only the following terms among the holomorphic soft terms: 
\dis{ \label{soft}
-{\cal L}_{\rm soft}\supset y_NA_N\widetilde{N}^ch_uH_d + m_B^2H_uH_d + {\rm h.c.} 
}
Since the SM gauge quantum numbers of $\{H_d,H_u\}$ are the same as $\{L,L^c\}$, 
the $y_N$ term of \eq{superPot} would be the same as the $y_N$ term of \eq{Vlep}, if other quantum numbers are ignored. 
Unlike the model in \eq{Vlep}, we introduce an extra  U(1)$_{Z^\prime}$ gauge symmetry, whose charge assignment is presented in Table \ref{tab:U(1)charge}.    
As seen above, of course, introduction of an extra non-Abelian gauge symmetry is very helpful for raising the radiative Higgs mass, avoiding the LP problem. 
In this paper, however, we attempt to raise it just  with an Abelian gauge symmetry, 
considering also the helps coming from the $\lambda$ term in \eq{superPot} and soft terms.  
As a result, only one pair of vector-like superfields is introduced. 
If necessary, one can extend the extra Abelian gauge symmetry U(1)$_{Z^\prime}$ to U(1)$_{Z1}\times$U(1)$_{Z2}\times$U(1)$_{Z3}\times\cdots$, 
which could much enhance the mechanism for evading LP    
without introducing more matter. 
Nonetheless, we will focus on the case only with one extra U(1) in this paper. 

\begin{table}[!h]
\begin{center}
\begin{tabular}
{c||c|c|c} 
Superfields &~ $H_u$, $N^c$, ($D^c$, $N_H^c$) ~&~ $H_d$, $N$, ($D$, $N_H$) ~&~ MSSM superfields,  $S$ 
 \\
 \hline
U(1)$^\prime$ charge ~& $q$ & $-q$ & $0$ 
\end{tabular}
\end{center}\caption{U(1)$_{Z^\prime}$ charge assignment. Only extra vector-like superfields carry  the U(1)$_{Z^\prime}$ charges of $\pm q$, while all the NMSSM superfields remain neutral.    
}\label{tab:U(1)charge}
\end{table}

Since the U(1)$_{Z^\prime}$ assigns the non-zero charges only to the extra vector-like superfields as shown in Table \ref{tab:U(1)charge}, 
it forbids mixing between the ordinary MSSM superfields and the newly introduced vector-like superfields in the bare superpotential. 
Accordingly, the new SU(2)$_L$ doublets $\{H_d,H_u\}$ don't induce unacceptable FCNC phenomena. 

$\{N^c,N\}$ can play the role of the Higgs for spontaneous breaking of U(1)$_{Z^\prime}$, if they got VEVs. 
Just for simplicity of the mass spectrum, 
one can introduce additional singlets, $\{N_H^c,N_H\}$ for breaking U(1)$_{Z^\prime}$. 
They can acquire VEVs just through the same mechanism with that for the MSSM Higgs. 
We don't discuss it here in details.   

In order to maintain the SM gauge coupling unification 
in the (N)MSSM, $\{H_d,H_u\}$ need to be supplemented 
with relatively heavier colored particles, 
$\{D^c,D\}$ to compose $\{{\bf 5},\overline{\bf 5}\}$ of SU(5) or $\{{\bf 5}_{-2},\overline{\bf 5}_{2}\}$ of flipped SU(5) \cite{flippedSU5}.  
They could eventually decay to SM fermions and neutral particles via, e.g. $W\supset q_iH_dD^c$, 
where $q_i$ is an MSSM quark doublet. 
In fact, relatively heavier extra colored particles $\{D^c,D\}$ can cure the small deviation of the gauge coupling unification appearing at the two-loop level  in the MSSM. 
In this paper, however, we don't discuss also it.  

Although we introduced $\{D^c,D;N_H^c,N_H\}$ for completeness of the model, they don't play an essential role for raising the radiative Higgs mass: 
$\{D^c,D\}$ could make just indirect contributions for improving the RG behaviors of various relevant Yukawa couplings, which will be discussed later.
Except for it, $\{D^c,D;N_H^c,N_H\}$ are almost spectators, concerning the radiative corrections of the Higgs which will be discussed below. 
Nonetheless, we note that many string models provide extra vector-like pairs of $\{H_d,H_u;D^c,D\}$ 
as well as singlets 
and extra U(1)s \cite{stringMSSM}. 
In this sense, introducing one pair of them and studying their phenomenological implications, 
in particular, the effects on the Higgs mass would be important, because the little hierarchy problem threatens the traditional status of the MSSM at the moment. 

\subsection{Mass spectrum}

In the bases of  $({H}_d^*(n_h^*),\widetilde{N}^* ;H_u(n_h^c),\widetilde{N}^c)$, 
the squared mass matrix for the neutral scalar fields in this model takes the following form:
\begin{eqnarray} \label{Mmatrix}
{\cal M}_{\cal B}^2 \approx \left[
\begin{array}{cc|cc}
|\mu_H|^2+|y_{N}h_u|^2 & \mu_{N}^*y_{N}h_u 
& m_B^2 & y_{N}A_Nh_u  \\
\mu_{N}y_{N}^{*}h_u^* & |\mu_{N}|^2 & 0 & 0
\\ \hline 
m_B^{2*} & 0 & |\mu_H|^2 & \mu_H^*y_{N}h_u  \\
y_{N}^*A_N^*h_u^* & 0 & \mu_Hy_{N}^*h_u^* & |\mu_{N}|^2+|y_{N}h_u|^2
\end{array}\right] + 
\left[
\begin{array}{c}
{\rm nonholomorphic}
\\
{\rm soft~mass~matrix}
\end{array}
\right], ~ 
\end{eqnarray}
where we ignored the contributions coming from  $y_N^\prime$, the MSSM $\mu$-term (or VEV of $S$), and ``$D$-term'' 
owing to their relative smallness. 
%
%
%
One can suppose that the ``nonholomorphic soft mass matrix'' takes a simple diagonal form: 
${\rm diag.}\left(\widetilde{m}_H^2,\widetilde{m}_N^2;\widetilde{m}_H^2,\widetilde{m}_N^2\right)$. 
Actually, the expressions of the mass eigenvalues for \eq{Mmatrix} are too complicate. 
Thus, we set 
$\widetilde{m}_H^2=\widetilde{m}_N^2\equiv \widetilde{m}^2$, 
assuming $|\widetilde{m}_H^2-\widetilde{m}_N^2|\equiv \Delta \widetilde{m}^2\ll \widetilde{m}^2$, i.e. SU(2)$_L$ doublets and singlets are almost degenerate as in the (s)top sector,   
and will study the following two limited cases for relatively simple analytic expressions:  
\begin{eqnarray}
\left\{
\begin{array}{l}
 \vspace{0.2cm}
~\widetilde{m}^2~,~~|A_N|^2~,~~|\mu_H|^2~\gg ~|m_B|^2  ~,~~|\mu_N|^2 ~,~~\Delta \widetilde{m}^2 
\qquad {\bf\rm Case~ I} ,
\\
~\widetilde{m}^2~,~~|A_N|^2~,~~|m_B|^2~\gg ~|\mu_H|^2  ~,~~|\mu_N|^2 ~,~~\Delta \widetilde{m}^2  
\qquad {\bf\rm Case~ II} .
\end{array}
\right.
\end{eqnarray} 
 
For Case I, then, the four eigenvalues of the squared mass matrix \eq{Mmatrix} are presented as   
\begin{eqnarray} \label{EvaluesI}
{\bf\rm Case~ I} \qquad \left\{
\begin{array}{l}
\vspace{0.2cm}
M_{B1,2}^2\approx \left(|\mu_H|^2+\widetilde{m}^2\right) + \alpha_\pm |y_Nh_u|^2 + \beta_\pm \frac{|y_Nh_u|^4}{|\mu_H|^2} ~, 
\\ \vspace{0.2cm}
M_{B3}^2\approx \widetilde{m}^2 -a_I^2|y_Nh_u|^2+a_I^2\left(2+a_I^2\right)\frac{|y_Nh_u|^4}{|\mu_H|^2} ~,
\\
M_{B4}^2\approx \widetilde{m}^2 ~,
\end{array}
\right. 
\end{eqnarray} 
where we expand the eigenvalues in powers of $|h_u|^2$ 
up to its quartic terms just for future convenience. 
In \eq{EvaluesI}, $a_I^2$ is defined as $a_I^2\equiv |A_N|^2/|\mu_H|^2$, and the coefficients, $\alpha_\pm$ and $\beta_\pm$ are given by 
\dis{
&\alpha_\pm \equiv 
1+\frac{a_I^2}{2}\pm\frac12\sqrt{a_I^2(4+a_I^2)} ~,
\\
&\beta_\pm \equiv 
 -\frac{a_I^2(2+a_I^2)}{2}\mp\frac{a_I^2\left[2+a_I^2(4+a_I^2)\right]}{2\sqrt{a_I^2(4+a_I^2)}} .
%
}
By setting $\widetilde{m}^2=a_I^2=0$ in \eq{EvaluesI}, we can obtain also the eigenvalues of the squared mass matrix for the fermionic fields, ${\cal M}_{\cal F}^2$:  
%
\dis{ \label{MF}
M_{F}^2=\left\{|\mu_H|^2+|y_Nh_u|^2,~|\mu_H|^2+|y_Nh_u|^2,~0,~0\right\} .  
}
%
%
For Case II, the eigenvalues of ${\cal M}_{\cal B}^2$ are expressed as follows:  
\begin{eqnarray} \label{EvaluesII}
{\bf\rm Case~ II} \quad \left\{
\begin{array}{l}
\vspace{0.2cm}
M_{B1,2}^2\approx \left(\widetilde{m}^2\pm |m_B|^2\right) + \gamma_\pm|y_Nh_u|^2 + \delta_\pm\frac{|y_Nh_u|^4}{|m_B|^2} ~, 
\\ \vspace{0.2cm}
M_{B3}^2\approx \widetilde{m}^2 +|y_Nh_u|^2-a_{II}^2\frac{|y_Nh_u|^4}{|m_B|^2} ~,
\\
M_{B4}^2\approx \widetilde{m}^2 ~,
\end{array}
\right. 
\end{eqnarray} 
where $a_{II}^2\equiv |A_N|^2/|m_B|^2$, and the coefficients, $\gamma_\pm$ and $\delta_\pm$ are 
\dis{
&\gamma_\pm \equiv 
\frac12\left(1\pm a_{II}^2\right) ,
\\
&\delta_\pm \equiv 
\frac{a_{II}^2}{2}\pm \frac18\left(1-a_{II}^4\right) .
}
Unless we assume  a non-vanishing VEV for the new Higgs doublets,
$\widetilde{m}^2$ should be greater than $|m_B|^2$ 
such that $M_{B2}^2>0$. 
Of course, the eigenvalues of ${\cal M}_{\cal F}^2$ in Case II are still given by \eq{MF}.

\subsection{Radiative Higgs potential} 

With the mass spectra of Eqs.~(\ref{EvaluesI}), (\ref{MF}), and (\ref{EvaluesII}), one can calculate 
the radiative corrections by $\{H_d, H_u; N^c, N\}$. 
Concerning the radiative Higgs mass and its renormalization, it is convenient to read them from the Coleman-Weinberg potential \cite{CW}: 
\dis{ \label{CWpot}
\Delta V=\frac{1}{32\pi^2}{\rm Tr}\bigg[{\cal M}_{\cal B}^4\left\{{\rm log}\frac{{\cal M}_{\cal B}^2}{Q^2}-\frac32\right\}
-{\cal M}_{\cal F}^4\left\{{\rm log}\frac{{\cal M}_{\cal F}^2}{Q^2}-\frac32\right\}\bigg] , 
}
where $Q$ denotes the renormalization scale. 
$\Delta V$ is expanded in powers of $|\delta h_u|^2$: 
($\equiv |h_u|^2-v_h^2{\rm sin}^2\beta$):  
\dis{ \label{expansion}
\Delta V\approx \Delta V_0+\left(\partial_{h_u}\partial_{h_u^*}\Delta V\right)_0 |\delta h_u|^2+\frac{1}{2!2!}\left(\partial^2_{h_u}\partial^2_{h_u^*}\Delta V\right)_0 |\delta h_u|^4+\cdots  .
}
For Case I, the coefficients of the quadratic 
and quartic terms of $\delta h_u$ in \eq{expansion}  are estimated as 
\begin{eqnarray} \label{quadI}
&&\left(\partial_{h_u}\partial_{h_u^*}\Delta V\right)_0\approx 
\frac{|y_N|^2}{8\pi^2}
\bigg[F_Q(|\mu_H|^2+\widetilde{m}^2)-F_Q(|\mu_H|^2) 
\\
&&\qquad\qquad \qquad\quad 
+ \frac{a_{I}^2}{2}
\left\{F_Q(|\mu_H|^2+\widetilde{m}^2)-F_Q(\widetilde{m}^2)\right\}\bigg]  ,
\nonumber 
\\
&&\left(\partial^2_{h_u}\partial^2_{h_u^*}\Delta V\right)_0\approx\frac{|y_N|^4}{4\pi^2}\bigg[
{\rm log}\left(1+z^2\right)
-\frac{a_{I}^4}{2}{\rm log}\left(\frac{1+z^2}{z^2}\right)
\label{quartI} \\
&&\qquad\qquad \qquad\quad  
+a_{I}^2(2+a_{I}^2)\left\{1-z^2{\rm log}\left(\frac{1+z^2}{z^2}\right)\right\}\bigg] ,
\nonumber  
\end{eqnarray}
where the function $F_Q$ is given again by $F_Q(m^2)= m^2\{{\rm log}(\frac{m^2}{Q^2})-1\}$ as in \eq{VlepRadi}, 
and the parameter $z$ is defined as $z^2\equiv\widetilde{m}^2/|\mu_H|^2$.   
Here we neglected   
$|y_N|^2v_h^2{\rm sin}^2\beta$ in $F_Q$s and the logarithmic functions in Eqs.~(\ref{quadI}) and (\ref{quartI}), 
because it is supposed to be quite smaller than $|\mu_H|^2$.  
%
%
For Case II, the coefficients of \eq{expansion} are given by 
\begin{eqnarray}  \label{quadII}
&&\left(\partial_{h_u}\partial_{h_u^*}\Delta V\right)_0\approx 
\frac{|y_N|^2}{32\pi^2}
\bigg[F_Q(\widetilde{m}^2+|m_B|^2)+F_Q(\widetilde{m}^2-|m_B|^2)+2F_Q(\widetilde{m}^2)-4F_Q(|\mu_H|^2)\bigg] 
\qquad
\\
&&\qquad\qquad\qquad\quad + a_{II}^2 \left\{
F_Q(\widetilde{m}^2+|m_B|^2)-F_Q(\widetilde{m}^2-|m_B|^2)\right\}\bigg]  ,
\nonumber \\
&&\left(\partial^2_{h_u}\partial^2_{h_u^*}\Delta V\right)_0\approx\frac{|y_N|^4}{32\pi^2}\bigg[
\left\{2+6a_{II}^2+(1+4a_{II}^2-a_{II}^4)\frac{z^2}{\zeta^2}\right\}{\rm log}\left(z^2+\zeta^2\right)
\label{quartII} \\
&& 
\qquad\qquad\qquad\quad 
+\left\{2-6a_{II}^2+(-1+4a_{II}^2+a_{II}^4)\frac{z^2}{\zeta^2}\right\}{\rm log}\left(z^2-\zeta^2\right)
\nonumber \\
&& \qquad\qquad\qquad\quad   
+\left(4-8a_{II}^2\frac{z^2}{\zeta^2}\right){\rm log}z^2
-2\left(1-a_{II}^4\right) \bigg] , 
\nonumber 
\end{eqnarray}
where $\zeta^2\equiv |m_B|^2/|\mu_H|^2$.

\subsubsection{Renormalization}

The quadratic term in \eq{expansion} ($\equiv \Delta \widetilde{m}_2^2(Q)|_{H}$) with the coefficient of  \eq{quadI}  or (\ref{quadII}) depends on the renormalization scale $Q$. 
It renormalizes the tree-level soft mass parameter of $h_u$ appearing in the MSSM Lagrangian, $\widetilde{m}_{2}^2(Q)$ together with the (s)top contribution $\Delta \widetilde{m}_2^2(Q)|_{\rm top}$:  
\dis{ \label{renorm}
\widetilde{m}_{2}^2(Q)+\Delta \widetilde{m}_2^2(Q)|_{\rm top} + \Delta \widetilde{m}_2^2(Q)|_{H} . 
}
%
%
Inserting the RG solution of $\widetilde{m}_2^2(Q)$ into \eq{renorm} yields the low energy value of $\widetilde{m}_2^2$ ($\equiv \widetilde{m}_2^2|_{\rm EW}$), 
replacing $Q$ in \eq{renorm} by a cut-off scale $\Lambda$ \cite{CQW}. 
%
%
The soft terms are regarded as being generated at the  messenger scale of SUSY breaking, since the soft terms would become non-local operators above the messenger scale.\footnote{ 
In the minimal SUGRA model, the messenger scale is assumed to be the GUT scale. 
Generically, however, the messenger scale is model-by-model different. 
We don't specify it in this paper.}
Thus, the messenger scale is adopted as the cut-off scale, and so we have   
\dis{ \label{EWm2}
\widetilde{m}_2^2|_{\rm EW}\approx \widetilde{m}_0^2+ \Delta \widetilde{m}_2^2|_{\rm top}
+\Delta \widetilde{m}_2^2|_H , 
}
where $\widetilde{m}_0^2$ stands for the value of $\widetilde{m}_2^2$ at the scale that it is generated, namely $\Lambda$. 
The (s)top contribution, $\Delta \widetilde{m}_2^2|_{\rm top}$ is presented as \cite{book} 
\begin{eqnarray} \label{mt}
\Delta \widetilde{m}_2^2|_{\rm top}&\approx&\frac{3|y_t|^2}{16\pi^2}
\Bigg[F_Q(\widetilde{m}_{tL}^2)
+F_Q(\widetilde{m}_{tR}^2)
-2F_Q(m_t^2)\Bigg]_{Q=\Lambda}
\\ \nonumber 
&& +\frac{3|y_t|^2}{16\pi^2}\frac{|A_t|^2}{\widetilde{m}_{tL}^2-\widetilde{m}_{tR}^2}
\Bigg[F_Q(\widetilde{m}_{tL}^2)-F_Q(\widetilde{m}_{tR}^2)\Bigg]_{Q=\Lambda} ,
\end{eqnarray}
where $\widetilde{m}_{tL}^2$ ($\widetilde{m}_{tR}^2$) denotes the soft mass squared of SU(2)$_L$ doublet (singlet) stop.\footnote{Here we assumed $\widetilde{m}_{tL,R}^2\gg |A_t|^2$, under which 
\eq{mt} is a good approximation. 
If $A_t$ is comparable to the stop masses, however, 
$\widetilde{m}_{tL,R}^2$ in \eq{mt} should be replaced by the mass eigenvalues, $\widetilde{m}_{t1,2}^2$ after mass matrix diagonalization for a precise expression.} 
This is the dominant radiative correction to $\widetilde{m}_2^2|_{\rm EW}$ in the MSSM. 
Note that $\Delta \widetilde{m}_2^2|_{\rm top}$ is regular at $\widetilde{m}_{tL}^2= \widetilde{m}_{tR}^2$.  
Moreover, it is quite insensitive to $\widetilde{m}_{tL}^2- \widetilde{m}_{tR}^2$. 
In the limit of $\widetilde{m}_{tL}^2= \widetilde{m}_{tR}^2\equiv\widetilde{m}_t^2$, \eq{mt} approaches to a much simple form: 
\dis{ \label{m2^2stop}
\Delta \widetilde{m}_2^2|_{\rm top}\longrightarrow \frac{3|y_t|^2}{8\pi^2}
\widetilde{m}_t^2 {\rm log}\frac{\widetilde{m}_t^2}{\Lambda^2}\times\left(1+ \frac{|A_t|^2}{2\widetilde{m}_t^2}\right) . 
}
For Case I, $\Delta \widetilde{m}_2^2|_H$ in \eq{EWm2} 
is given by 
\begin{eqnarray} 
\Delta \widetilde{m}_2^2|_I&\approx&\frac{|y_N|^2}{8\pi^2}
\bigg[F_Q(|\mu_H|^2+\widetilde{m}^2)-F_Q(|\mu_H|^2)
\nonumber \\    
&& 
\quad+\frac{a_{I}^2}{2}\left\{F_Q(|\mu_H|^2+\widetilde{m}^2)-F_Q(\widetilde{m}^2)\right\}\bigg]_{Q=\Lambda}
\label{FI} \\ 
\nonumber 
&\longrightarrow& \frac{|y_N|^2}{8\pi^2}
\widetilde{m}^2 {\rm log}\frac{\widetilde{m}^2}{\Lambda^2}\times\left(1+\frac{a_{I}^2}{2z^2}\right)
 \qquad {\rm for}~~\Lambda^2~\gg~\widetilde{m}^2~\gg~ |\mu_H|^2 ,
\end{eqnarray}
while $\Delta \widetilde{m}_2^2|_H$ for Case II is  
\begin{eqnarray} 
\Delta \widetilde{m}_2^2|_{II}&\approx&\frac{|y_N|^2}{32\pi^2}
\bigg[F_Q(\widetilde{m}^2+|m_B|^2)+F_Q(\widetilde{m}^2-|m_B|^2)+2F_Q(\widetilde{m}^2)-4F_Q(|\mu_H|^2)
\nonumber  \\   
&& 
\qquad ~~ +a_{II}^2\left\{F_Q(\widetilde{m}^2+|m_B|^2)-F_Q(\widetilde{m}^2-|m_B|^2)\right\}\bigg]_{Q=\Lambda}
\label{FII} \\  
\nonumber 
&\longrightarrow& \frac{|y_N|^2}{8\pi^2}
\widetilde{m}^2 {\rm log}\frac{\widetilde{m}^2}{\Lambda^2}\times\left(1+\frac{a_{II}^2\zeta^2}{2z^2}\right)
 \qquad {\rm for}~~\Lambda^2~\gg~\widetilde{m}^2,~ |m_B|^2 ~\gg~|\mu_H|^2 . ~~
\end{eqnarray}
In the last lines of Eqs.~(\ref{FI}) and (\ref{FII}), 
we took the large $\Lambda$ limit.

As mentioned in Introduction, the EW scale or the Higgs VEV is determined by $\widetilde{m}_2^2|_{\rm EW}$ of \eq{EWm2} and other (soft) mass parameters:    
$\widetilde{m}_2^2|_{\rm EW}$ eventually participates in 
the one of the extremum conditions for the Higgs potential ($\partial_{h_u}V_H=\partial_{h_d}V_H=0$)  \cite{book,twoloop}:
\dis{ \label{extremum}
\widetilde{m}_2^2|_{\rm EW}+|\mu|_{\rm EW}^2 
\approx m_3^2|_{\rm EW}~{\rm cot}\beta 
+\frac{M_Z^2}{2}{\rm cos}2\beta ,
}
which $m_3^2$ denotes the ``$B\mu$ term'' coefficient. 
It should, of course, be fulfilled around the vacuum state. 
If $\widetilde{m}_2^2|_{\rm EW}$ is too large, it gives rise to a serious fine-tuning problem, 
because $\widetilde{m}_2^2|_{\rm EW}$ should be matched to the $Z$ boson mass squared, $M_Z^2$ [$=(g_2^2+g_Y^2)v_h^2\approx(91~ {\rm GeV})^2$] in \eq{extremum}. 
%
For naturalness of the EW scale and its perturbative stability, thus, the dimensionful parameters in Eqs.~(\ref{m2^2stop}), (\ref{FI}), and (\ref{FII}) 
should be small enough. 
Also, a lower mediation scale of SUSY breaking is  very helpful for relaxing the fine-tuning.


%
%
%
In the MSSM, \eq{m2^2stop} with stop mass heavier than $700~{\rm GeV}$ would make the biggest contributions to $\widetilde{m}_2^2|_{\rm EW}$. 
Moreover, $|A_t|^2/\widetilde{m}_t^2$ in \eq{m2^2stop} deteriorates the fine-tuning problem. 
In order to minimize the fine-tuning, thus, the stop mass needs to be as light as possible, and the $|A_t|$ should also be suppressed.  
With a stop mass much heavier than $700~{\rm GeV}$ and an $|A_t|$ comparable to it, however, 
the observed Higgs mass would be more easily explained, as will be seen later,  
even if the fine-tuning problem becomes worse.  
In this paper, we will take the experimental lower bound, $\widetilde{m}_{tL}^2\approx  \widetilde{m}_{tR}^2\equiv\widetilde{m}_t^2\approx (700~{\rm GeV})^2$,  
assuming a quite small $A_t$ term. 
Under this condition, we will attempt to account for the observed 126 GeV Higgs mass, utilizing other ingredients contained in this model.


Concerning the fine-tuning problem, 
smaller SUSY breaking $A$-term and SUSY(-breaking) masses would be required also for the extra vector-like fields $\{H_d,H_u;N^c,N\}$. 
Actually, light extra leptonic particles are still experimentally acceptable, only if they can immediately decay to the neutral particles.   
On the contrary, masses of extra colored particles are severely constrained from LHC data, and 
heavy enough extra colored particles coupled to the MSSM Higgs would cause a fine-tuning in the Higgs sector.    
It is the reason why we are particularly interested in the extra {\it colorless} particles.

In the limit of $\widetilde{m}^2\gg |\mu_H|^2$ 
and $\widetilde{m}^2\gg |m_B|^2$, however, 
$F_Q(|\mu_H|^2+\widetilde{m}^2)-F_Q(\widetilde{m}^2)$
and $F_Q(\widetilde{m}^2+|m_B|^2)-F_Q(\widetilde{m}^2-|m_B|^2)$ in Eqs.~(\ref{FI}) and (\ref{FII}) vanish, respectively. 
In this limit, therefore, larger values of $a_{I,II}^2$ can be taken without making the fine-tuning worse.
Instead, $|y_N|^2$ should be small enough, because the other terms of Eqs.~(\ref{FI}) and (\ref{FII}) increase in this case.  
In the limit of $\widetilde{m}^2\ll |\mu_H|^2$, 
in contrast, $F_Q(|\mu_H|^2+\widetilde{m}^2)-F_Q(\widetilde{m}^2)$ in Eq.~(\ref{FI}) becomes larger, while the first two terms of Eq.~(\ref{FI}) cancel each other. 
In this case, thus, $|y_N|^2|a_I|^2$ should be small enough for avoiding a too serious fine-tuning.
Note that $\widetilde{m}^2$ cannot be smaller than $|m_B|^2$ in Case II, if the extra Higgs don't get a non-zero VEV.  

In the case of \eq{Vlep}, the $A_N$-term was not considered. Instead, a large value of $y_N$ ($\gtrsim 1$) was possible \cite{Vlep}. So $\widetilde{m}^2$ could not be much greater than $|\mu_L|^2$. 
In this model, however, we can have a relatively large $A_N$-term. So $\widetilde{m}^2\gtrsim |\mu_H|^2$ is permitted, only if $|y_N|\lesssim 1$. 
For instance, if the U(1)$_{Z^\prime}$ gauge sector plays also the role of the SUSY breaking messenger \cite{Zprime}, or  the extra vector-like superfields $\{H_d,H_u;N^c,N\}$ carry also other gauge charges associated with a U(1)$^\prime$ mediation of SUSY breaking apart from U(1)$_{Z^\prime}$, 
a relatively large $A_N$ (and also $m_B^2$) term  can be generated, leaving intact the $A_t$ term.

\subsubsection{Radiative Higgs mass}

The quartic term in \eq{expansion} with the coefficient of Eq.~(\ref{quartI}) or (\ref{quartII}),   
which is independent of the renormalization scale $Q$  \cite{book}, 
makes contribution to the radiative correction to the {\it physical Higgs mass} together with the (s)top. 
Thus, the summation of all the tree-level and the radiative squared masses should yield the experimental value of the Higgs squared mass: 
\dis{ \label{fullmass}
m_h^2\approx M_Z^2{\rm cos^22\beta}+\lambda^2v_h^2{\rm sin}^22\beta +\Delta m_h^2|_{\rm top} + \Delta m_h^2|_{H} ~\approx ~ (126~{\rm GeV})^2 . 
}
Here the first and second terms are the tree-level Higgs mass of the MSSM and NMSSM, 
while the last two terms  correspond to the radiative corrections to it.  
The (s)top contribution $\Delta m_h^2|_{\rm top}$ in \eq{fullmass} is presented as \cite{book}  
\begin{eqnarray} \label{MSSMradiMass}
\Delta m_h^2|_{\rm top}&\approx& \frac{3v_h^2{\rm sin}^4\beta}{4\pi^2}|y_t|^4 
\Bigg[{\rm log}\left(\frac{\sqrt{\widetilde{m}_{tL}^2
\widetilde{m}_{tR}^2}}{m_t^2}\right) 
+\frac{|A_t|^2}{\widetilde{m}_{tL}^2-\widetilde{m}_{tR}^2} ~{\rm log}\left(\frac{\widetilde{m}_{tL}^2}{\widetilde{m}_{tR}^2}\right)
\\ \nonumber 
&&\qquad\qquad +\frac{|A_t|^4}{(\widetilde{m}_{tL}^2-\widetilde{m}_{tR}^2)^2}
\left\{1-\frac12\frac{\widetilde{m}_{tL}^2
+\widetilde{m}_{tR}^2}{\widetilde{m}_{tL}^2-\widetilde{m}_{tR}^2}
~{\rm log}\left(\frac{\widetilde{m}_{tL}^2}{\widetilde{m}_{tR}^2}\right)\right\}\Bigg] ,
\end{eqnarray}
which is regular also at $\widetilde{m}_{tL}^2= \widetilde{m}_{tR}^2$. In the limit of $\widetilde{m}_{tL}^2= \widetilde{m}_{tR}^2$ ($\equiv\widetilde{m}_t^2$), 
it approaches to 
\dis{ \label{MSSMradi}
\Delta m_h^2|_{\rm top}
\longrightarrow 
\frac{3v_h^2{\rm sin}^4\beta}{4\pi^2}|y_t|^4 
\Bigg[{\rm log}\left(\frac{\widetilde{m}_t^2}{m_t^2}\right)
+\frac{|A_t|^2}{\widetilde{m}_t^2}\left(1-\frac{1}{12}\frac{|A_t|^2}{\widetilde{m}_t^2}\right)\Bigg] . 
}
Note that $v_h^2{\rm sin}^4\beta|y_t|^4$ in \eq{MSSMradi} is simply written as $m_t^4/v_h^2$ as seen in \eq{TopStop}. 
Even if \eq{MSSMradi} is derived under the limit of $\widetilde{m}_{tL}^2= \widetilde{m}_{tR}^2$, 
it is approximately valid over the large  parameter space of $(\widetilde{m}_{tL}^2 ,\widetilde{m}_{tR}^2)$, 
unless they are extremely hierarchical.
Particularly, if $|A_t|^2>2(\widetilde{m}_{tL}^2+ \widetilde{m}_{tR}^2)$, \eq{MSSMradi} is almost the maximum that \eq{MSSMradiMass} is able to reach 
using $\widetilde{m}_{tL}^2$ and $\widetilde{m}_{tR}^2$ for a given $|A_t|^2/(\widetilde{m}_{tL}^2+\widetilde{m}_{tR}^2)$. 
Were it not for the last two terms in \eq{MSSMradi}, thus, $\widetilde{m}_t^2$ was only a useful parameter for raising the Higgs mass in the MSSM.  
Since the radiative Higgs mass is a logarithmic function of $\widetilde{m}_t^2$ in such a case, 
raising the Higgs mass using $\widetilde{m}_t^2$ 
would be a quite inefficient way. 
The last two terms in \eq{MSSMradi} are maximized when $|A_t|^2/\widetilde{m}_t^2=6$. 
As mentioned before, however, they make the fine-tuning problem more serious. 
In this paper, thus, we will discuss the radiative Higgs mass without considering the $A_t$ terms in \eq{MSSMradi}, as mentioned above.

The contribution to the radiative Higgs mass by $\{H_d, H_u; N^c, N\}$ in \eq{fullmass}, $\Delta m_h^2|_H$ is computed with the quartic coefficient of \eq{expansion}, i.e. Eq.~(\ref{quartI}) or (\ref{quartII}):  
\begin{eqnarray} \label{massI}
&&\Delta m_h^2|_{I}\approx \frac{v_h^2{\rm sin}^4\beta}{4\pi^2}|y_N|^4 \Bigg[
{\rm log}\left(1+z^2\right)
-\frac{a_{I}^4}{2}{\rm log}\left(\frac{1+z^2}{z^2}\right)
\nonumber \\ 
&&\qquad +a_{I}^2(2+a_{I}^2)\left\{1-z^2{\rm log}\left(\frac{1+z^2}{z^2}\right)\right\}\Bigg]
\equiv \frac{v_h^2{\rm sin}^4\beta}{4\pi^2}|y_N|^4 \times Z_I   
\\ 
&&\left(\longrightarrow\frac{v_h^2{\rm sin}^4\beta}{4\pi^2}|y_N|^4\left[{\rm log}z^2+\frac{a_I^2}{z^2}\left(1-\frac{1}{12}\frac{a_I^2}{z^2}\right)\right]\quad {\rm for}~z^2,~a_I^2\gg1\right) 
\nonumber 
\end{eqnarray}
and
\begin{eqnarray} \label{massII}
&&\Delta m_h^2|_{II}\approx \frac{v_h^2{\rm sin}^4\beta}{32\pi^2}|y_N|^4\bigg[
\left\{2+6a_{II}^2+(1+4a_{II}^2-a_{II}^4)\frac{z^2}{\zeta^2}\right\}{\rm log}\left(z^2+\zeta^2\right)
\nonumber \\
&&\qquad
+\left\{2-6a_{II}^2+(-1+4a_{II}^2+a_{II}^4)\frac{z^2}{\zeta^2}\right\}{\rm log}\left(z^2-\zeta^2\right)
\\
&&\qquad +\left(4-8a_{II}^2\frac{z^2}{\zeta^2}\right){\rm log}z^2
-2\left(1-a_{II}^4\right) \bigg]
\equiv \frac{v_h^2{\rm sin}^4\beta}{4\pi^2}|y_N|^4 \times Z_{II} 
\nonumber 
\\ 
&&\left(\longrightarrow\frac{v_h^2{\rm sin}^4\beta}{4\pi^2}|y_N|^4\left[{\rm log}z^2+a_{II}^2\frac{\zeta^2}{z^2}\left(1-\frac{a_{II}^2}{12}\frac{\zeta^2}{z^2}\right)\right]\quad {\rm for}~z^2/\zeta^2 ,~a_{II}^2\gg1\right) 
\nonumber 
\end{eqnarray}
for Case I and II, respectively. 
Although we don't necessarily require $z^2,~a_{I}^2\gg 1$ or $z^2/\zeta^2,~a_{II}^2\gg1$, we presented the asymptotic expressions of $\Delta m_h^2|_{I,II}$ in Eqs.~(\ref{massI}) and (\ref{massII}) just for comparison with \eq{MSSMradi}. 
The shapes of $Z_I$ and $Z_{II}$ defined in Eqs.~(\ref{massI}) and (\ref{massII}) are displayed in Fig. 1 and 2 in terms of $(z,a_{I})$ and $(z,a_{II})$, respectively. 
As seen in Fig. 1-(a) and 2-(a), the $A_N$-term in \eq{soft} is quite helpful for raising the radiative Higgs mass:  
$Z_{I}$ and $Z_{II}$, which are proportional to the radiative Higgs mass, rapidly grow along the $z$ ($a_{I}$ and $a_{II}$) direction(s) for relatively larger $a_{I}$ and $a_{II}$ (a smaller $z$). 
We note that $Z_{I}$ and $Z_{II}$ are in the range of 
\dis{
0 ~\lesssim ~ Z_{I,II} ~\lesssim ~ 4.5
}
for $1\lesssim z\lesssim 5$ and $0\lesssim a_I\lesssim 5$ in Case I, 
and for $\zeta\approx 2.5$, $2.5\lesssim z\lesssim 5$, and $0\lesssim a_{II}\lesssim 2$ in Case II.  
These parameter ranges are translated into the following scopes in terms of the Lagrangian parameters: 
\begin{eqnarray}
\left\{
\begin{array}{l}
\vspace{0.2cm}
(200~{\rm GeV})^2\lesssim \widetilde{m}^2\lesssim (1~{\rm TeV})^2 , 
\\
\vspace{0.2cm}
\qquad 0\lesssim A_N\lesssim 1~{\rm TeV} 
\\
{\rm for}~~\mu_H=200~{\rm GeV} ~\gg ~|m_B^2|^{1/2}
\end{array}
\right.
~~
\left\{
\begin{array}{l}
\vspace{0.2cm}
(400~{\rm GeV})^2\lesssim \widetilde{m}^2\lesssim (2~{\rm TeV})^2 , 
\\ \vspace{0.2cm}
\qquad 0\lesssim A_N\lesssim 2~{\rm TeV} 
\\
{\rm for}~~\mu_H=400~{\rm GeV} ~\gg ~|m_B^2|^{1/2}
\end{array}
\right.
\quad {\rm in~Case ~I} , ~~
\end{eqnarray}   
and  
\begin{eqnarray}
\left\{
\begin{array}{l}
\vspace{0.2cm}
(500~{\rm GeV})^2\lesssim \widetilde{m}^2\lesssim (1~{\rm TeV})^2 , 
\\ \vspace{0.2cm}
\qquad 0\lesssim A_{N}\lesssim 1~{\rm TeV}
\\ \vspace{0.2cm}
{\rm for}~~\mu_H=200~{\rm GeV} 
~~{\rm and}~~
\\
\qquad M_B^2=(500~{\rm GeV})^2
\end{array}
\right.
~~~
\left\{
\begin{array}{l}
\vspace{0.2cm}
(750~{\rm GeV})^2\lesssim \widetilde{m}^2\lesssim (1.5~{\rm TeV})^2 , 
\\ \vspace{0.2cm}
\qquad 0\lesssim A_{N}\lesssim 1.5~{\rm TeV}
\\ \vspace{0.2cm}
{\rm for}~~\mu_H=300~{\rm GeV} 
~~{\rm and}~~
\\
\qquad M_B^2=(750~{\rm GeV})^2
\end{array}
\right.
~~ {\rm in~Case ~II} .~~
\end{eqnarray}
They easily pass the LEP constraint on extra leptons \cite{JSW,PDG}. 
Moreover, the extra charged leptons rapidly decay to the neutral ones and SM fermions in this model.

\begin{figure}
\begin{center}
\subfigure[]
{\includegraphics[width=0.52\linewidth]{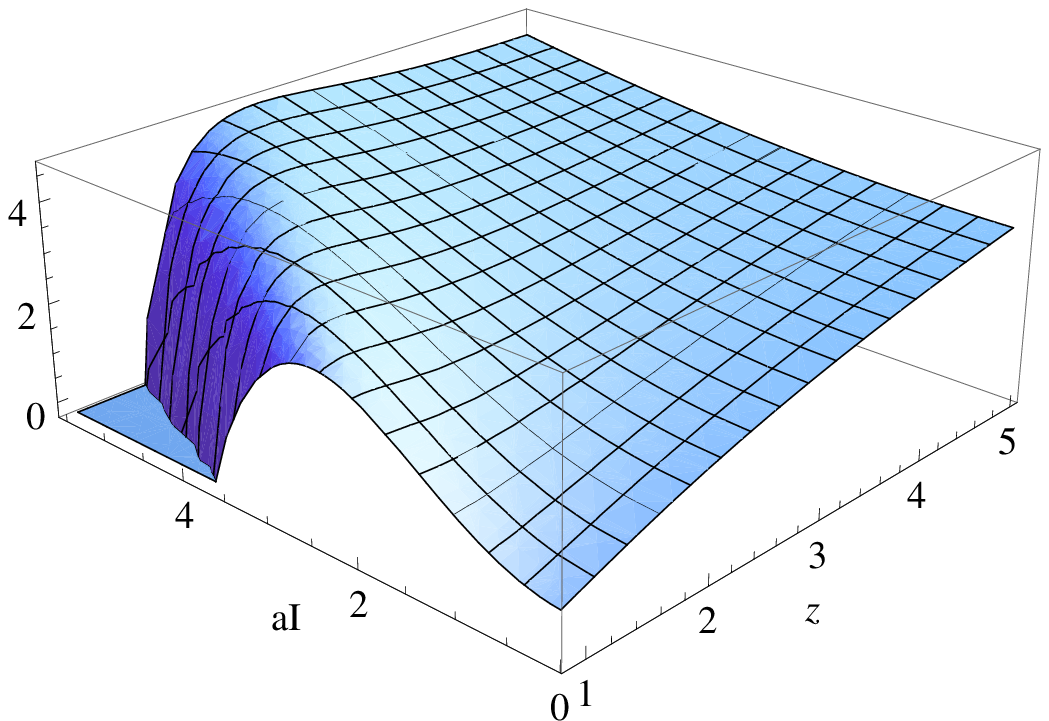}}
\hspace{0.2cm}
\subfigure[] 
{\includegraphics[width=0.4\linewidth]{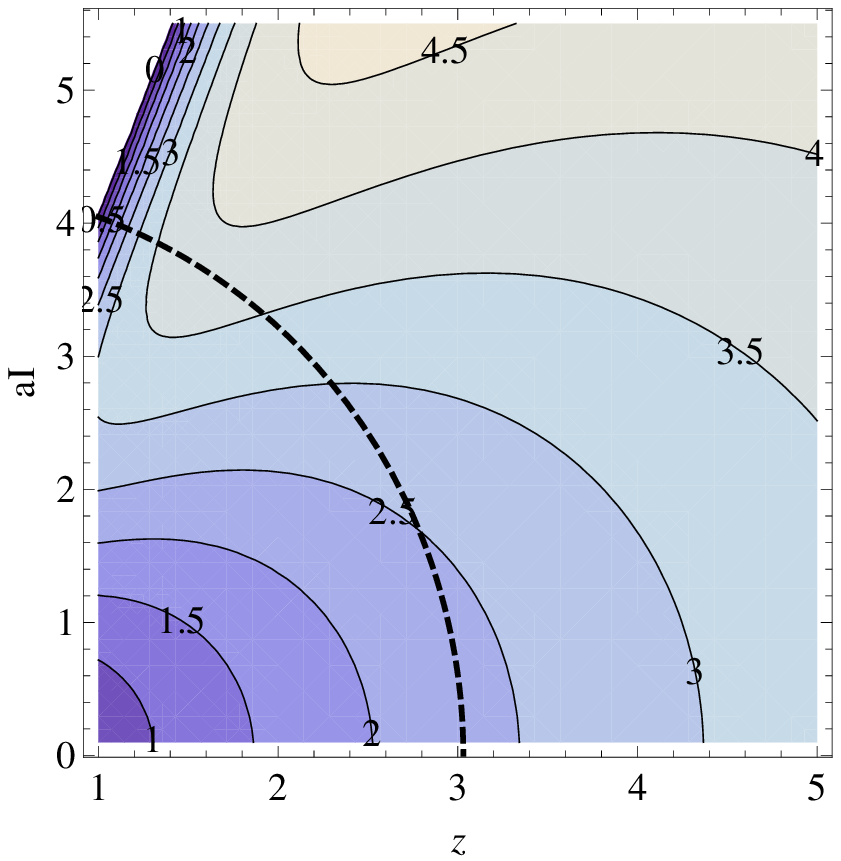}}
\end{center}
\caption{(a) 3D plot of $Z_I$. It is proportional to the radiative Higgs mass by $\{H_d, N^c\}$ in Case I. 
(b) Contour plot of $Z_I$. The parameter space of $(z,a_{I})$ inside the dotted line could avoid a serious fine-tuning of the Higgs sector for $\mu_H\approx 400~{\rm GeV}$,  while the entire parameter space can do for $\mu_H\approx 200~{\rm GeV}$. 
}
\end{figure}

\begin{figure}
\begin{center}
\subfigure[]
{\includegraphics[width=0.52\linewidth]{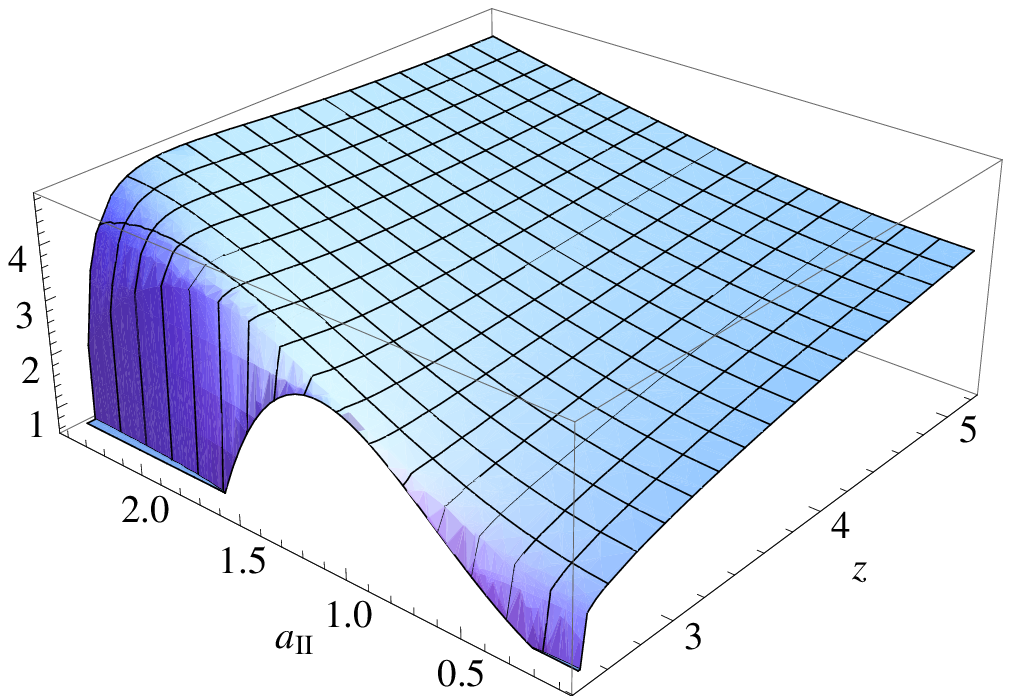}}
\hspace{0.2cm}
\subfigure[] 
{\includegraphics[width=0.4\linewidth]{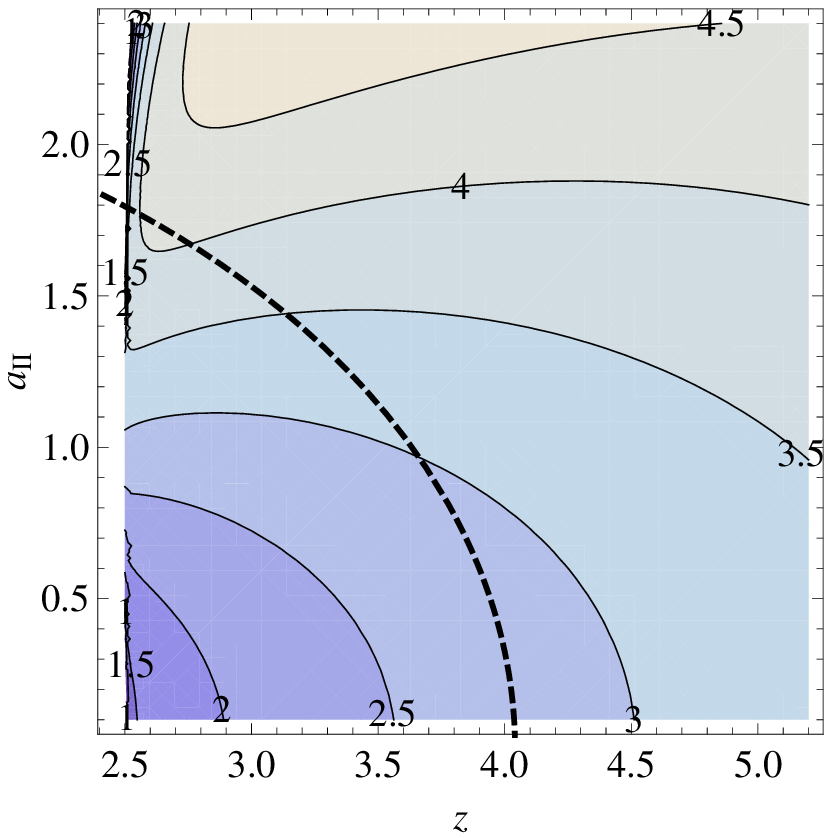}}
\end{center}
\caption{(a) 3D plot of $Z_{II}$. It is proportional to the radiative Higgs mass by $\{H_d, N^c\}$ in Case II. 
(b) Contour plot of $Z_{II}$. The parameter space of $(z,a_{II})$ inside the dotted line could avoid a serious fine-tuning of the Higgs sector for $\mu_H\approx 300~{\rm GeV}$,  while the entire parameter space can do for $\mu_H\approx 200~{\rm GeV}$.    
}
\end{figure}

The above ranges of parameters don't much affect the oblique parameters, $T$ and $S$. 
If $\mu_H=\mu_N$ and $y_N^\prime=0$, $\Delta T$ and $\Delta S$ by the new vector-like fermions are estimated as \cite{extramatt2}
\dis{
\Delta T= 0.54~  y_N^4 {\rm sin}^4\beta
\left(\frac{100~{\rm GeV}}{\mu_H}\right)^2 ,
\qquad
\Delta S= 0.13~  y_N^2 {\rm sin}^2\beta
\left(\frac{100~{\rm GeV}}{\mu_H}\right)^2 .
}
The experimental best fit requires that $0.01\lesssim \Delta S\lesssim 0.17$ ($1\sigma$) for $\Delta T\approx 0.12$ \cite{ST}.  
It can be satisfied only if $\mu_H\gtrsim 153~{\rm GeV}$, $119~{\rm GeV}$, $89~{\rm GeV}$ for $y_N=0.85, ~0.75, ~0.65$, respectively, and ${\rm tan}\beta=50$. 
Even if $\mu_N$ is much smaller than $y_N\langle h_u\rangle$, the results are not much different: 
it can be shown that this case yields a bit smaller $(\Delta T, \Delta S)$ than those for $\mu_H=\mu_N$ using the formulas of Refs.~\cite{extramatt2,oblique}. 
The contributions of the new scalars to $(\Delta T, \Delta S)$ would be more suppressed due to their heavy masses.

\subsection{126 GeV Higgs boson vs. Fine-tuning}

The Higgs mass in this model is determined with Eqs.~(\ref{fullmass}), (\ref{MSSMradi}), (\ref{massI}), and (\ref{massII}). 
If the $A_t$ terms are ignored in \eq{MSSMradi}, 
they are recast into 
\dis{ \label{126Higgs}
|y_N|^4Z_{I,II}=\frac{4\pi^2}{(v_hs^2\beta)^2}\left[
m_h^2-M_Z^2(1-2s^2\beta)^2
-4\lambda^2v_h^2 s^2\beta(1-s^2\beta)
-\frac{3m_t^4}{4\pi^2v_h^2}
~{\rm log}\left(\frac{\widetilde{m}_t^2}{m_t^2}\right)\right]
}
where $m_h^2\approx (126~{\rm GeV})^2$ and $s^2\beta\equiv {\rm tan}^2\beta/(1+{\rm tan}^2\beta)$. 
$Z_I$ and $Z_{II}$ are defined in Eqs.~(\ref{massI}) and (\ref{massII}), respectively. 
Only if the above equation is satisfied, thus, 
the observed 126 GeV Higgs mass is explained.  
However, we will constrain the parameters in \eq{126Higgs} such that the newly introduced superfields $\{H_d, H_u; N^c, N\}$ 
don't raise a more serious fine-tuning problem in the Higgs sector 
than that already resulted from the (s)top. 
From Eqs.~(\ref{m2^2stop}), (\ref{FI}), and (\ref{FII}), thus, we have 
\dis{ \label{FT} 
z^2+\frac12 a_{I}^2 &~\lesssim~ 3\frac{|y_t|^2}{|y_N|^2}\frac{\widetilde{m}_t^2}{|\mu_H|^2} 
\qquad~~ {\rm in~~Case~I}
\\
z^2+\frac{\zeta^2}{2} a_{II}^2 &~\lesssim~ 3\frac{|y_t|^2}{|y_N|^2}\frac{\widetilde{m}_t^2}{|\mu_H|^2} 
\qquad~~ {\rm in~~Case~II} 
}
for the least tuning.
Here we assumed $\Lambda^2\gg \widetilde{m}_t^2,~\widetilde{m}^2, ~|\mu_H|^2, ~|m_B|^2$. 
%
%
Only with the experimental lower bound of the stop, $m_t^2\approx (700~{\rm GeV})^2$,  
we will show that the Higgs boson mass of 126 GeV can still be explained with our extension of the NMSSM, \eq{superPot} 
in the parameter space \eq{FT}, as mentioned above.

In order to meet \eq{126Higgs}, a larger value of $y_N$ is favored: were it not for the $A_N$- and $\lambda$ terms, $|y_N|$ should be greater than unity.  
For a larger parameter space of \eq{FT}, on the other hand, a smaller $y_N$ is preferred.
Note that the {\it quartic} power of $|y_N|$ appears in \eq{126Higgs}, while just its {\it quadratic} power dependence arises in \eq{FT}.  
In Ref.~\cite{Vlep}, thus, $|y_N|\gtrsim 1$ was taken. 
Instead, a relatively smaller $\widetilde{m}^2$ and $|\mu_H|^2$ compensate it.   
In this case, a unwanted LP would appear below the GUT scale. 
Because of that, an extra non-Abelian gauge symmetry,  SU(2)$_{Z^\prime}$ was introduced in Ref.~\cite{Vlep}.
In this paper, we will consider the possibility of $|y_N|\lesssim 1$ and an extra U(1)$_{Z^\prime}$ gauge symmetry. 
Instead, we take into account of the $\lambda$ and $A_N$ terms in Eqs.~(\ref{superPot}) and (\ref{soft}). 
As seen in Fig. 1 and 2, a relatively larger value of $A_N$
can efficiently raise $Z_{I,II}$ or the radiative Higgs mass.  

For $|\mu_H|\approx 200~{\rm GeV}$, the entire parameter spaces of $(z,a_{I})$ and $(z,a_{II})$ in Fig. 1 and 2 are well-inside the upper bounds of \eq{FT}. 
For $|\mu_H|\approx 400~{\rm GeV}$ [$300~{\rm GeV}$] in Case I [II], however, only the space inside the dotted line in Fig. 1-(b) [2-(b)] fulfills \eq{FT}.  
In both cases, we set $\widetilde{m}_t^2 = (700~{\rm GeV})^2$. 

%

\section{Landau-pole constraint} \label{sec:LP}

As seen in Eqs.~(\ref{massI}), (\ref{massII}), and (\ref{126Higgs}), larger values of $\lambda$ and $y_N$ are preferred for raising the tree-level and radiative Higgs masses. 
If such Yukawa coupling constants are too large, however, they would blow up below the GUT scale.  
In this section, we investigate the maximally allowed 
values for them at low energy scale ($\lesssim 1~{\rm TeV}$) for avoiding the LP constraints. 
  
From \eq{superPot} and the ordinary superpotential of the NMSSM, the anomalous dimensions of the relevant superfields can read as follows: 
\begin{eqnarray} \label{gammaExtra}
{\rm Extra ~Fields}~~~~&&\left\{
\begin{array}{l}
16\pi^2\gamma^{H_d}_{H_d}=|y_N|^2
-\frac32g_2^2-\frac{3}{10}g_1^2-2q^2g_{Z^\prime}^2 ,
\vspace{0.2cm} \\ 
16\pi^2\gamma^{N^c}_{N^c}=2|y_N|^2
-2q^2g_{Z^\prime}^2 ,
\vspace{0.2cm} 
\\ 
16\pi^2\gamma^{S}_{S}~=2|\lambda|^2 ,
\end{array}
\right.
\\ \label{gammaHiggs}
{\rm MSSM ~Higgs}~~~~&&\left\{
\begin{array}{l}
16\pi^2\gamma^{h_u}_{h_u}=2|y_N|^2 + |\lambda|^2
+3|y_t|^2-\frac32 g_2^2-\frac{3}{10}g_1^2 ,
\vspace{0.2cm} \\
16\pi^2\gamma^{h_d}_{h_d}=|\lambda|^2 + 3|y_b|^2
+|y_\tau|^2-\frac32 g_2^2-\frac{3}{10}g_1^2 ,
\end{array}
\right.
\\ \label{gammaMatt}
{\rm MSSM ~Matter}~~&&\left\{
\begin{array}{l}
16\pi^2\gamma^{q_3}_{q_3}=|y_t|^2 + |y_b|^2
-\frac83g_3^2 -\frac32g_2^2-\frac{1}{30}g_1^2 ,
\vspace{0.2cm} \\
16\pi^2\gamma^{u^c_3}_{u^c_3}=2|y_t|^2 
-\frac83g_3^2 -\frac{8}{15}g_1^2 ,
\vspace{0.2cm} \\
16\pi^2\gamma^{d^c_3}_{d^c_3}=2|y_b|^2  
-\frac83g_3^2 -\frac{2}{15}g_1^2 ,
\vspace{0.2cm} \\
16\pi^2\gamma^{l_3}_{l_3}=|y_\tau|^2 
-\frac32g_2^2 -\frac{3}{10}g_1^2 ,
\vspace{0.2cm} \\
16\pi^2\gamma^{e^c_3}_{e^c_3}=2|y_\tau|^2 
-\frac{6}{5}g_1^2 . 
\end{array}
\right. 
\end{eqnarray}
Here we considered only the third generation of the MSSM matter, $\{q_3, u^c_3, d^c_3; l_3, e^c_3\}$, concerning the MSSM Yukawa couplings: 
$y_t$, $y_b$, and $y_\tau$ denote the top quark, bottom quark, and tau's Yukawa couplings, respectively. 
The $g_{3,2,1}^2$ in Eqs.~(\ref{gammaExtra}), (\ref{gammaHiggs}), and  (\ref{gammaMatt}) stand for the three MSSM gauge couplings 
associated with SU(3)$_c$, SU(2)$_L$, and U(1)$_Y$ gauge interactions.  
As discussed before, we introduced an extra U(1)$_{Z^\prime}$ gauge symmetry in order to resolve the LP problem associated with $y_N$. 
$g_{Z^{\prime}}^2$ terms appear in \eq{gammaExtra} due to the U(1)$_{Z^\prime}$ gauge interactions with the charge assignment in Table \ref{tab:U(1)charge}. 
Such MSSM and extra gauge interactions make the {\it negative} contributions to the anomalous dimensions.
Considering the relevant superpotential, 
one can readily write down the RG equations for the Yukawa coupling constants: 
\begin{eqnarray} 
\label{RGeq}
\left\{
\begin{array}{l}
\frac{d|y_N|^2}{dt}= \frac{|y_N|^2}{8\pi^2}\bigg[4|y_N|^2+|\lambda|^2+3|y_t|^2-3g_2^2-\frac35g_1^2-4q^2g_{Z^\prime}^2
\bigg] ,
\vspace{0.3cm}\\
\frac{d|\lambda|^2}{dt}= \frac{|\lambda|^2}{8\pi^2}\bigg[|y_N|^2+4|\lambda|^2+3|y_t|^2
+3|y_b|^2+|y_\tau|^2-3g_2^2-\frac35g_1^2
\bigg] ,
\vspace{0.3cm} \\
\frac{d|y_t|^2}{dt}=\frac{|y_t|^2}{8\pi^2}\bigg[|y_N|^2+|\lambda|^2+6|y_t|^2+|y_b|^2-\frac{16}{3}g_3^2-3g_2^2-\frac{13}{15}g_1^2
\bigg] ,
\vspace{0.3cm} \\
\frac{d|y_b|^2}{dt}=\frac{|y_b|^2}{8\pi^2}\bigg[|\lambda|^2+|y_t|^2+6|y_b|^2+|y_\tau|^2-\frac{16}{3}g_3^2-3g_2^2-\frac{7}{15}g_1^2
\bigg] ,
\vspace{0.3cm} \\
\frac{d|y_\tau|^2}{dt}=\frac{|y_\tau|^2}{8\pi^2}\bigg[|\lambda|^2+3|y_b|^2+4|y_\tau|^2-3g_2^2-\frac{9}{5}g_1^2
\bigg] ,
\end{array}
\right. 
\end{eqnarray}
where $t$ parametrizes the renormalization scale, 
$t-t_0={\rm log}(Q/M_{\rm GUT})$.

The one-loop RG equations for the three MSSM gauge couplings are integrable. 
The RG solutions to them take the following form: 
\begin{eqnarray} \label{RGMSSM}
g_k^2(t)=\frac{g_U^2}
{1+\frac{g_U^2}{8\pi^2}b_k(t_0-t)} \qquad {\rm for} ~~k=3,~2,~1 ,
\end{eqnarray}
where $b_k$ ($k=3,2,1$) denotes the beta function coefficients for the gauge couplings of SU(3)$_c$, SU(2)$_L$ and U(1)$_Y$ [with the SU(5) normalization]. 
%
In the presence of the one pair of $\{{\bf 5},\overline{\bf 5}\}$ as in our case, 
they are given by $b_k=(-2,2,38/5)$,  
and the unified gauge coupling $g_U^2$ at the GUT scale is estimated as $0.62$.

The solution to the RG equation of the extra U(1)$_{Z^\prime}$ gauge coupling has also a similar form to \eq{RGMSSM}: 
\begin{eqnarray}  \label{Z'gauge}
g_{Z^\prime}^2(t)=\frac{g_{U}^2}
{1+\frac{g_{U}^2}{8\pi^2}b_{Z^\prime}(t_0-t)} ~~ {\rm for}~~ t>t_{Z^\prime}   
\end{eqnarray}
where $t_{Z^\prime}$ parametrizes the U(1)$_{Z^\prime}$ breaking scale $M_{Z^\prime}$ 
[$t_{Z^\prime}-t_0\equiv {\rm log}(M_{Z^\prime}/M_{\rm GUT})$].
U(1)$_{Z^\prime}$ can be spontaneously broken at TeV scale in the same manner of SU(2)$_L\times$U(1)$_Y$ breaking mechanism in the MSSM.
$b_{Z^\prime}$ in \eq{Z'gauge} denotes the beta function coefficient of $g^2_{Z^\prime}$, which is model-dependent.
If the extra colored particles $\{D,D^c\}$ and the 
U(1)$_{Z^\prime}$ breaking Higgs $\{N_H^c,N_H\}$ carrying U(1)$_{Z^\prime}$ of $\pm q$ are also included,  
the charge assignment in Table \ref{tab:U(1)charge} 
yields $b_{Z^\prime}=14q^2$. 
We assume that all the gauge couplings are unified, $g_3^2=g_2^2=g_1^2=g_{Z\prime}^2\equiv g_U^2$ at the $M_G$ scale. 

From the first equation in \eq{RGeq}, we can expect that the LP constraint can be relaxed by the additional negative contributions coming from the $g_{Z^\prime}^2$ terms. 
As a result, the allowed maximal values for $y_{N}$ 
at low energies can be lifted up, compared to the case that the U(1)$_{Z^\prime}$ gauge symmetry is absent, and so the radiative Higgs mass can be raised with a larger $y_N$, particularly for a large ${\rm tan}\beta$ as seen in Eqs.~(\ref{massI}) and (\ref{massII}). 
Consequently, even relatively smaller value of $\lambda$ can be consistent with the observed 126 GeV Higgs mass: the lower bound of $\lambda$ and the upper bound of ${\rm tan}\beta$ could be relaxed, compared to the case of NMSSM, \eq{NMSSMpara}. 

Now let us discuss the cases of {\bf A.} $y_N=0$ and ${\lambda\neq 0}$, {\bf B.} $y_N\neq 0$ and $\lambda=0$, {\bf C.} $y_N\neq 0$ and $\lambda\neq 0$  without the U(1)$_{Z^\prime}$ gauge symmetry, and {\bf D.} $y_N\neq 0$ and $\lambda\neq 0$ with the  U(1)$_{Z^\prime}$ gauge symmetry in order.  

\subsection{$y_N=0$, $\lambda\neq 0$} 

If the $y_N$ coupling is not introduced in the superpotential \eq{superPot}, 
U(1)$_{Z^\prime}$ cannot affect the radiative Higgs potential, because only the extra vector-like superfields have the non-zero charge of it, 
as seen in Table \ref{tab:U(1)charge}.  
Nonetheless, we have introduced extra one pair of $\{{\bf 5},\overline{\bf 5}\}$, and so the MSSM gauge couplings become a bit larger by them at higher energy scales, compared to those of the original (N)MSSM. 
Hence, the LP constraint of $\lambda$ can slightly be relaxed by the stronger MSSM gauge interactions via the RG equation of $\lambda$ of \eq{RGeq} \cite{Masip}, even if no $g_{Z^\prime}^2$ contribution is there.  
Since Yukawa couplings monotonically increase with energy, throughout this paper
we require that all the squared Yukawa coupling constants discussed here should be smaller than the perturbativity bound, $4\pi\approx 12$ at the GUT scale.

\begin{table}[!h]
\begin{center}
\begin{tabular}
{c|c|c|c|c|c}
  &   ~${\rm tan}\beta=2$~  &  ~${\rm tan}\beta=3$~ & ~${\rm tan}\beta=4$~ & ~${\rm tan}\beta=5$~  & ~${\rm tan}\beta=5.5$   
  \\ \hline
$\lambda_{\rm max}$  & $0.69$ & $0.74$ & $0.75$ & $0.76$ & $0.76$   
\\
$\lambda_{\rm 126}$    &  $0.58$  &  $0.61$ & $0.67$ & $0.73$ & $0.77$  
\end{tabular}
\end{center}\caption{Maximally allowed low energy values of $\lambda$ ($=\lambda_{\rm max}$) and 
$\lambda$s needed for explaining the 126 GeV Higgs mass ($=\lambda_{126}$) when $\widetilde{m}_t=700~{\rm GeV}$ and $y_N=0$. $\lambda_{126}$ should be smaller than $\lambda_{\rm max}$ for evading a LP below the GUT scale. 
}\label{tab:y_N=0}
\end{table}

Table \ref{tab:y_N=0} lists  
the maximally allowed low energy values of 
$\lambda$ ($\equiv \lambda_{\rm max}$), avoiding the LP below the GUT scale, and 
the needed values of $\lambda$ for explaining 126 GeV Higgs mass ($\equiv \lambda_{126}$) 
with $\widetilde{m}_t=700 ~{\rm GeV}$, depending on ${\rm tan}\beta$, 
{\it when the $y_N$ coupling is absent, i.e. for $y_N=0$}.
$\lambda_{\rm max}$ can be estimated using the RG equations \eq{RGeq}, while $ \lambda_{126}$ is determined by \eq{126Higgs} setting $y_{N}=0$.
Since $\lambda_{126}$ cannot exceed $\lambda_{\rm max}$ to avoid the LP below the GUT scale, ${\rm tan}\beta$ should be smaller than 5.5 in this case.  
We see that even if there is no $y_N$ coupling in the superpotential, the allowed ranges of the values for 
${\rm tan}\beta$ and $\lambda$ become quite wider  
in the presence of one extra pair of $\{{\bf 5},\overline{\bf 5}\}$:
\dis{
0.58\leq \lambda < 0.77
\qquad 
{\rm for}~~ 2\leq {\rm tan}\beta < 5.5.
}
In this case, still only relatively small values of ${\rm tan}\beta$ are consistent with the observed Higgs mass and the perturbativity of the model up to the GUT scale.

\subsection{$y_N\neq 0$, $\lambda=0$} 

Now, we discuss the LP constraint of $y_N$, when $\lambda=0$ in the superpotential \eq{superPot}.  
Since $H_d$ and $N^c$ are charged under U(1)$_{Z^\prime}$, 
the U(1)$_{Z^\prime}$ gauge interaction is helpful for relaxing the LP problem for $y_N$.  
If the U(1)$_{Z^\prime}$ gauge symmetry is not introduced, however, 
there does not appear the $g_{Z^\prime}^2$ term in the RG equation of $y_N$ in \eq{RGeq}. 
As a result, elusion of a LP for $y_N$ is not efficient. 
For comparison, Table \ref{tab:lambda=0} displays the results of the both cases, $q=0$ and $q=\sqrt{5}$.  
$y_{N{\rm max}}$ means the maximally allowed low energy value of $y_N$ needed for evading LP, 
which is obtained by performing the analysis of the 
RG equations \eq{RGeq}. 
$Z_{I,II{\rm min}}$ is the value of $Z_{I}$ or $Z_{II}$ required to account for the 126 GeV Higgs mass for each $y_{N{\rm max}}$.  
It can be estimated with \eq{126Higgs}.
$y_{N{\rm min}}$ indicates the value of $y_N$ 
needed for explaining the 126 GeV Higgs mass 
when $Z_{I,II}$ takes the maximum value, $Z_{I,II}=4.5$. 
For perturbativity up to the GUT scale and naturalness of the model, 
$y_N$ should be located between $y_{N{\rm max}}$ and $y_{N{\rm min}}$.

\begin{table}[!h]
\begin{center}
\begin{tabular}
{c||c|c|c|c|c|c|c|c}
${q=0}$    &   ${\rm tan}\beta=4$  &  ${\rm tan}\beta=6$ & ${\rm tan}\beta=8$ & ${\rm tan}\beta=10$  & ${\rm tan}\beta=20$ &  ${\rm tan}\beta=30$ & ${\rm tan}\beta=40$ & ${\rm tan}\beta=50$   
  \\ \hline
$y_{N{\rm max}}$  & $0.75$ & $0.76$ & $0.77$ & $0.77$ & $0.77$  & $0.76$ & $0.76$ & $0.75$  
\\
$Z_{I,II{\rm min}}$    &  $13.71$  &  $8.32$ & $6.59$ & $5.80$ & $4.77$  & $4.68$ & $4.73$  & $4.99$
\\
\hline\hline
${q=\sqrt{5}}$    &   ${\rm tan}\beta=4$  &  ${\rm tan}\beta=6$ & ${\rm tan}\beta=8$ & ${\rm tan}\beta=10$  & ${\rm tan}\beta=20$ &  ${\rm tan}\beta=30$ & ${\rm tan}\beta=40$ & ${\rm tan}\beta=50$   
  \\ \hline
$y_{N{\rm max}}$  & $0.85$ & $0.86$ & $0.86$ & $0.86$ & $0.86$  & $0.86$ & $0.85$ & $0.84$  
\\
$y_{N{\rm min}}$  & $-$ & $-$ & $0.84$ & $0.82$ & $0.78$  & $0.77$ & $0.77$ & $0.77$  
\\
$Z_{I,II{\rm min}}$    &  $8.65$  &  $5.28$ & $4.16$ & $3.67$ & $3.03$  & $2.95$ & $2.98$  & $3.12$
\end{tabular}
\end{center}\caption{Maximally allowed low energy values of $y_N$ ($=y_{N\rm max}$) and $Z_{I}$ or $Z_{II}$ needed for explaining the 126 GeV Higgs mass ($=Z_{I,II{\rm min}}$) for each $y_{N{\rm max}}$ when $\widetilde{m}_t=700~{\rm GeV}$. 
$y_{N{\rm min}}$s indicate the values of $y_N$ needed for the 126 GeV Higgs mass with $Z_{I,II}=4.5$.
The first two lines list the results in the absence of the gauged U(1)$_{Z^\prime}$, 
while the last three lines are those in the presence of the U(1)$_{Z^\prime}$ gauge symmetry with $q=\sqrt{5}$, and $b_{Z^\prime} =14q^2$. 
}\label{tab:lambda=0}
\end{table}

In Table \ref{tab:lambda=0}, the first two lines of $(y_{N{\rm max}},Z_{I,II{\rm max}})$  correspond to the results of $q=0$ case, while the last three lines of $(y_{N{\rm max}},y_{N{\rm min}},Z_{I,II{\rm max}})$ 
to the case of $q=\sqrt{5}$.   
In the absence of the gauged U(1)$_{Z^\prime}$ i.e. 
in the $q=0$ case, $Z_{I,II{\rm min}}$s exceed 4.5 throughout the range of $2\lesssim {\rm tan}\beta\lesssim 50$.     
Hence, if one takes $0\lesssim Z_{I,II}\lesssim 4.5$
to ameliorate the fine-tuning problem in the Higgs sector, $y_N$ should be greater than $y_{N{\rm max}}$
for the observed 126 GeV Higgs mass,  
and so it diverges below the GUT scale. 
On the contrary, if the U(1)$_{Z^\prime}$ gauge interaction is turned on, it is possible to elude the fine-tuning and LP problems, explaining the observed Higgs mass in the large ${\rm tan}\beta$ range,  
\dis{
7 \lesssim {\rm tan}\beta \lesssim 50 ,
}
and $0.77\lesssim y_N\lesssim 0.86$ depending on ${\rm tan}\beta$. 
For ${\rm tan}\beta\lesssim 7$, however, $y_{N{\rm min}}$ exceeds $y_{N{\rm max}}$. 

\subsection{$y_N\neq 0$, $\lambda\neq 0$ without a gauged U(1)$_{Z^\prime}$}

Table \ref{tab:GlobalU(1)} shows the results of the case, in which both the $\lambda$ and $y_N$ couplings in \eq{superPot} are turned on, but the U(1)$_{Z^\prime}$ is not gauged [or $q=0$ in \eq{RGeq}].  
$y_{N{\rm max}}$ is defined as the maximal value of $y_N$ at low energy ($\lesssim 1~{\rm TeV}$) such that {\it all} the Yukawa couplings considered here, $(\lambda, y_N, y_t, y_b, y_\tau)$ do not blow up below the GUT scale for a given low energy value of $\lambda$. 
Because of the LP constraint, $y_N$ could not be large enough in this case. 
Accordingly, $Z_{I}$ and $Z_{II}$ should be excessively larger than 4.5 in most parameter space as shown in Table \ref{tab:GlobalU(1)} in order to account for the 126 GeV Higgs mass. 
It means that $z^2$ (or $\widetilde{m}^2$) and $a_{I,II}$ (or $A_N$) need to be quite large,  violating \eq{FT}.

\begin{table}[!h]
\begin{center}
\begin{tabular}
{c||ccc||ccc||ccc||ccc} 
  &     & ${\rm tan}\beta=2$  &   &  & ${\rm tan}\beta=4$  &    & & ${\rm tan}\beta=6$ & & & ${\rm tan}\beta=10$ & 
  \\
\hline 
$\lambda$  & $0.5$ & $0.6$ & $0.7$ & $0.5$ & {\bf $\framebox[1.3\width]{0.6}$} & $0.7$ & $0.5$ & $0.6$ & $0.7$ & $0.5$ & $0.6$ & $0.7$ 
\\
\hline 
$y_{N{\rm max}}$  & $0.64$ & $0.59$ & $\times$ & $0.72$ & $0.69$ & $0.57$ & $0.72$ & $0.70$ & $0.60$ & $0.73$ & $0.71$ & $0.61$
\\
$Z_{I,II{\rm min}}$  & $19.82$ & $-8.39$ & $\times$ & $7.51$ & $3.87$ & $-3.83$ & $6.41$ & $5.17$ & $5.18$ & $5.64$ & $5.68$ & $8.83$
\\ \hline\hline
  &     & ${\rm tan}\beta=20$  &   &  & ${\rm tan}\beta=30$  &    & & ${\rm tan}\beta=40$ & & & ${\rm tan}\beta=50$ & 
  \\
\hline 
$\lambda$  & $0.0$ & $0.2$ & $0.4$ & $0.0$ & $0.2$ & $0.4$ & $0.0$ & $0.2$ & $0.4$ & $0.0$ & $0.2$ & $0.4$ 
\\
\hline 
$y_{N{\rm max}}$  & $0.77$ & $0.76$ & $0.75$ & $0.76$ & $0.76$ & $0.74$ & $0.76$ & $0.75$ & $0.73$ & $0.75$ & $0.74$ & $0.71$
\\
$Z_{I,II{\rm min}}$  & $4.77$ & $4.85$ & $5.15$ & $4.68$ & $4.78$ & $5.19$ & $4.73$ & $4.85$ & $5.45$ & $4.99$ & $5.17$ & $6.14$
\end{tabular}
\end{center}\caption{Maximally allowed values of $y_N$ for given $\lambda$s at low energy and the corresponding minimal values of $Z_{I}$ or $Z_{II}$ consistent with the 126 GeV Higgs mass for $\widetilde{m}_t=700~{\rm GeV}$ and $q=0$.
}\label{tab:GlobalU(1)}
\end{table}

Although we present only the results for $\lambda =0.5$, $0.6$, $0.7$ for ${\rm tan}\beta=2,4,6,10$ in Table \ref{tab:GlobalU(1)}, 
$Z_{I,II}$ turn out to be much larger than $4.5$ for $0\leq\lambda < 0.5$ and $2\leq {\rm tan}\beta\leq 10$.   
For ${\rm tan}\beta=20,30,40,50$, $Z_{I,II{\rm min}}$ monotonically increases with $\lambda$. 
However, $Z_{I,II{\rm min}}$ already exceeds 4.5 even for $\lambda=0$. 
Note that the results for $\lambda=0$ when ${\rm tan}\beta=20,30,40,50$
reproduce the corresponding results of $q=0$ in Table \ref{tab:lambda=0}.

However, the $y_N$ coupling is definitely helpful for raising the radiative Higgs mass, and so the consistent parameter space in this case should become broader.  
We note that the sign of $Z_{I,II}$ should flip between $0.5$ ($0.6$) and $0.6$ ($0.7$) for ${\rm tan}\beta=2$ ($4$). 
Hence $Z_{I,II}=0$ at some point between them.  
$Z_{I,II}=0$ is effectively equivalent to the case of $y_N=0$ ($<y_{N{\rm max}}$, of course) discussed in Table \ref{tab:y_N=0} at low energy in explaining the Higgs mass, 
because the left hand side of \eq{126Higgs} vanish anyway in the both cases, 
$|y_{N}|^4Z_{I,II}=0$.   
Comparing with $\lambda_{126}$s of Table \ref{tab:y_N=0}, thus,  
we can expect that $|y_N|^4Z_{I,II}=0$ at $\lambda=0.58$ (0.67) for ${\rm tan}\beta=2$ ($4$). 
Since now we have the $y_N$ coupling and $A_N$ term, 
which are helpful for explaining the observed Higgs mass, even $\lambda$ smaller than $\lambda_{126}$ can be permitted.  
Actually, the following parameter ranges meet the perturbative constraint up to the GUT scale, 
explaining the observed Higgs mass:     
\begin{eqnarray} \label{GlobalU(1)_lambda}
\left\{
\begin{array}{l}
0.56 < \lambda < 0.58 \qquad {\rm for}~~{\rm tan}\beta=2 ,
\\
0.57 < \lambda < 0.61 \qquad {\rm for}~~{\rm tan}\beta=3 ,
\\
0.59 < \lambda < 0.67 \qquad {\rm for}~~{\rm tan}\beta=4 ,
\\
0.61 < \lambda < 0.73 \qquad {\rm for}~~{\rm tan}\beta=5 ,
\end{array}
\right.
\end{eqnarray}
which means that the consistent {\it points} in Table \ref{tab:y_N=0} become narrow {\it bands} in the parameter space, when the $y_N$ coupling and $A_N$ term are introduced.  
Around the lower bounds of $\lambda$ in \eq{GlobalU(1)_lambda}, it turns out that $y_N$ should be restricted 
to $0.61$, $0.68$, $0.69$, $0.70$ for ${\rm tan}\beta=2,3,4,5$, respectively, since the upper bound coming from the LP constraint and the lower bound for the 126 GeV Higgs mass get to merge together. 
On the contrary, around the upper bounds of $\lambda$ in \eq{GlobalU(1)_lambda}, $y_N$ is constrained only by LP, because the 126 GeV Higgs mass has been already explained with the the maximal $\lambda$s and so 
the radiative Higgs mass correction by $\{H_d,N^c\}$ proportional to $|y_N|^4 Z_{I,II}$ 
should vanish at low energy. 
Thus, we have just the (trivial) constraints, $y_N(\approx 0) < y_{N{\rm max}} = 0.61, ~0.67, ~0.65, ~0.45$ for ${\rm tan}\beta=2,3,4,5$, respectively, around the upper bound of $\lambda$.

\subsection{$y_N\neq 0$, $\lambda\neq 0$ with the gauged U(1)$_{Z^\prime}$}

Table \ref{tab:LocalU(1)} presents the results when not only the $\lambda$, $y_N$ terms, but also the U(1)$_{Z^\prime}$ gauge symmetry are introduced particularly with $q=\sqrt{5}$. 
Since the U(1)$_{Z^\prime}$ gauge coupling, $g_{Z^\prime}$ monotonically increase with energy and $q^2g_{U}^2/4\pi\approx 0.25$, the perturbativity 
of U(1)$_{Z^\prime}$ gauge interaction is guaranteed throughout the energy range from the EW to the GUT scale.  
Then, the negative contribution by U(1)$_{Z^\prime}$ gauge interactions could make the LP constraint on $y_N$ remarkably relaxed, as mentioned above. 
Although one takes a more larger value of $q$, e.g. $q=\sqrt{20}$, which would be almost the maximal value of $q$ to maintain the perturbativity of U(1)$_{Z^\prime}$ gauge interaction at the GUT scale, 
it turns out that a conspicuous improvement of the allowed parameter space is not achieved.   
%

In Table \ref{tab:LocalU(1)}, $y_{N{\rm max}}$ again indicates the maximally allowed value of $y_N$ for a given $\lambda$ at low energy: 
only if $y_N$ is smaller than $y_{N{\rm max}}$ around 1 TeV energy scale, any Yukawa couplings considered here do not reach the perturbativity bound ($\lambda^2, y_N^2, y_{t,b,\tau}^2<4\pi\approx 12$) below the GUT scale. 
As in the previous tables, 
$Z_{I,II{\rm min}}$ stands for the value of $Z_{I}$ or $Z_{II}$ 
required for explaining the 126 GeV Higgs mass, when the corresponding $y_{N{\rm max}}$ is taken. 
$y_{N{\rm min}}$ means the value of $y_{N}$
needed for explaining the observed Higgs mass when  $Z_{I,II}=4.5$. 

In Table \ref{tab:LocalU(1)}, the $\lambda$s satisfying both the LP and Higgs mass constraints are 
written inside the boxes.  
We note that 
the allowed range of $\lambda$ is remarkably enlarged  
particularly for larger values of ${\rm tan}\beta \gtrsim 8$. 
[Actually, the results of ${\rm tan}\beta=8$ show a similar pattern to the case of ${\rm tan}\beta=10$, even if they are not displayed in Table \ref{tab:LocalU(1)}.]
Thus, we have 
\begin{eqnarray} 
\left\{
\begin{array}{l}
0.5\lesssim \lambda \lesssim 0.6 ~~\qquad\quad\qquad {\rm for} ~~{\rm tan}\beta = 4 ,
\\
0.4\lesssim \lambda \lesssim 0.7 ~~\qquad\quad\qquad {\rm for} ~~{\rm tan}\beta = 6 ,
\\
0 \leq \lambda \lesssim 0.6,~0.5, ~0.4, \qquad {\rm for} ~~{\rm tan}\beta = \{8,~10,~20,~30\},~40,~50,  
\end{array}
\right.
\end{eqnarray}
respectively, and roughly $0.75\lesssim y_N\lesssim 0.85$ depending on ${\rm tan}\beta$. 
Note that the lists of $\lambda=0$ are coincident with the results of $q=\sqrt{5}$ in Table \ref{tab:lambda=0}. 
Comparing with the parameter range of the NMSSM, \eq{NMSSMpara}, much larger values of ${\rm tan}\beta$ are also allowed, and the lower bound of $\lambda$ is remarkably relieved. 
In particular, the lower bound of $\lambda$ disappears for ${\rm tan}\beta\gtrsim 8$. 
It is because the $y_N$ and $A_N$ terms of  $\{H_d,N^c\}$ significantly raise the radiative Higgs mass particularly for large ${\rm tan}\beta$. 
Moreover, the U(1)$_{Z^\prime}$ gauge interaction 
makes it possible that their contributions are further enhanced. 

We note that the sign of $Z_{I,II{\rm min}}$ is flipped between $\lambda=0.6$ and $0.7$ for ${\rm tan}\beta=4$ in Table \ref{tab:LocalU(1)}. 
One can expect that $Z_{I,II{\rm min}}$ vanishes at a point between them, which is effectively equivalent to $y_N=0$ of Table \ref{tab:y_N=0} in explaining the Higgs mass, because $|y_N|^4Z_{I,II}=0$ in the both cases. 
By performing a similar analysis to \eq{GlobalU(1)_lambda}, we get the following results:
\begin{eqnarray} \label{LocalU(1)_lambda}
\left\{
\begin{array}{l}
0.55 < \lambda < 0.58 \qquad {\rm for}~~{\rm tan}\beta=2 ,
\\
0.53 < \lambda < 0.61 \qquad {\rm for}~~{\rm tan}\beta=3 ,
\\
0.51 < \lambda < 0.67 \qquad {\rm for}~~{\rm tan}\beta=4 ,
\\
0.46 < \lambda < 0.73 \qquad {\rm for}~~{\rm tan}\beta=5 ,
\end{array}
\right.
\end{eqnarray}
which are wider than the former results in  \eq{GlobalU(1)_lambda}, 
because of the gauged U(1)$_{Z^\prime}$. 
Around the lower bounds of $\lambda$ in \eq{LocalU(1)_lambda}, $y_N$ should be restricted 
to $0.70$, $0.78$, $0.81$, $0.82$ for ${\rm tan}\beta=2,3,4,5$, respectively, and $Z_{I,II}\approx 4.5$, while $y_N=0$ around the upper bounds.  
The upper bounds of \eq{LocalU(1)_lambda} should coincide with those of \eq{GlobalU(1)_lambda} and also 
the results of Table \ref{tab:y_N=0}.
Between the lower and upper bounds of $\lambda$, 
sizable intervals of $y_N$ can be allowed: 
e.g. for ${\rm tan}\beta=4$ and $\lambda=0.6$ ($0.55$), 
the permitted range of $y_N$ is given by $0.66$ $(0.75)\lesssim y_N\lesssim 0.78$ ($0.80$), 
as seen from $y_{N{\rm max}}$ and $y_{N{\rm min}}$ of  $\lambda=0.6$ 
in Table  \ref{tab:LocalU(1)}. 
For ${\rm tan}\beta=5$ and $\lambda=0.65$ ($0.55$), 
the allowed range of $y_N$ turns out to be $0.63$ $(0.76)\lesssim y_N\lesssim 0.76$ ($0.80$). 
For the smaller ${\rm tan}\beta$, the intervals are relatively narrower. 
However, they all should rapidly shrink to $y_N=0$ around the upper bounds of $\lambda$ in \eq{LocalU(1)_lambda}, which correspond to the original NMSSM limit at low energy.
%

%
%
\begin{table}[!h]
\begin{center}
\begin{tabular}
{c||ccccccccc} 
  &     &   &   &  & ${\rm tan}\beta=4$  &    & &  & 
  \\
\hline 
$\lambda$  & $0.0$ & $0.1$ & $0.2$ & ~$0.3$ & $0.4$ & {\bf $\framebox[1.3\width]{0.5}$} & ~{\bf $\framebox[1.3\width]{0.6}$} & $0.7$ & $0.75$  
\\
\hline 
$y_{N{\rm max}}$  & ~~$0.85$~~ & ~~$0.85$~~ & ~~$0.84$~~ & ~~$0.83$ & $0.82$ & $0.81$ & ~~~$0.78$~ & ~~$0.61$~~ & ~~$0.20$~~ 
\\
$y_{N{\rm min}}$  & ~~$-$~~ & ~~$-$~~ & ~~$-$~~ & ~~$-$ & $-$ & $0.81$ & ~~~$0.66$~ & ~~$-$~~ & ~~$-$~~ 
\\
$Z_{I,II{\rm min}}$  & $8.65$ & $8.50$ & $8.07$ & ~~$7.32$ & $6.22$ & $4.63$ & ~~$2.36$ & $-2.92$ & $\times$ 
\\
\hline\hline
  &     &   &   &  & ${\rm tan}\beta=6$  &    & &  & 
  \\
\hline 
$\lambda$  & $0.0$ & $0.1$ & $0.2$ & ~$0.3$ & {\bf $\framebox[1.3\width]{0.4}$} & {\bf $\framebox[1.3\width]{0.5}$} & ~{\bf $\framebox[1.3\width]{0.6}$} & {\bf $\framebox[1.3\width]{0.7}$} & $0.75$  
\\
\hline 
$y_{N{\rm max}}$  & ~~$0.86$~~ & ~~$0.85$~~ & ~~$0.85$~~ & ~~$0.84$ & $0.83$ & $0.82$ & ~~~$0.79$~ & ~~$0.65$~~ & ~~$0.32$~~ 
\\
$y_{N{\rm min}}$  & ~~$-$~~ & ~~$-$~~ & ~~$-$~~ & ~~$-$ & $0.83$ & $0.79$ & ~~~$0.73$~ & ~~$0.62$~~ & ~~$-$~~ 
\\
$Z_{I,II{\rm min}}$  & $5.28$ & $5.22$ & $5.07$ & ~~$4.80$ & $4.40$ & $3.85$ & ~~$3.14$ & $3.76$ & $33.66$ 
\\
\hline\hline
  &     &   &   &  & ${\rm tan}\beta=10$  &    & &  & 
  \\
\hline 
$\lambda$  & {\bf $\framebox[1.3\width]{0.0}$} & {\bf $\framebox[1.3\width]{0.1}$} & {\bf $\framebox[1.3\width]{0.2}$} & ~{\bf $\framebox[1.3\width]{0.3}$} & {\bf $\framebox[1.3\width]{0.4}$} & {\bf $\framebox[1.3\width]{0.5}$} & ~{\bf $\framebox[1.3\width]{0.6}$} & $0.7$ & $0.75$  
\\
\hline 
$y_{N{\rm max}}$  & ~~$0.86$~~ & ~~$0.86$~~ & ~~$0.86$~~ & ~~$0.85$ & $0.84$ & $0.82$ & ~~$0.80$~ & ~~$0.66$~~ & ~~$0.33$~~ 
\\
$y_{N{\rm min}}$  & ~~$0.82$~~ & ~~$0.81$~~ & ~~$0.81$~~ & ~~$0.80$ & $0.79$ & $0.77$ & ~~$0.75$~ & ~~$-$~~ & ~~$-$~~ 
\\
$Z_{I,II{\rm min}}$  & $3.67$ & $3.65$ & $3.62$ & ~~$3.59$ & $3.54$ & $3.54$ & ~$3.49$ & $6.45$ & $93.48$ 
\\
\hline\hline
  &     &   &   &  & ${\rm tan}\beta=20$  &    & &  & 
  \\
\hline 
$\lambda$  & {\bf $\framebox[1.3\width]{0.0}$} & {\bf $\framebox[1.3\width]{0.1}$} & {\bf $\framebox[1.3\width]{0.2}$} & ~{\bf $\framebox[1.3\width]{0.3}$} & {\bf $\framebox[1.3\width]{0.4}$} & {\bf $\framebox[1.3\width]{0.5}$} & ~{\bf $\framebox[1.3\width]{0.6}$} & $0.7$ & $0.75$  
\\
\hline 
$y_{N{\rm max}}$  & ~~$0.86$~~ & ~~$0.86$~~ & ~~$0.85$~~ & ~~$0.85$ & $0.84$ & $0.82$ & ~~$0.79$~ & ~~$0.60$~~ & ~~$\times$~~ 
\\
$y_{N{\rm min}}$  & ~~$0.78$~~ & ~~$0.78$~~ & ~~$0.78$~~ & ~~$0.77$ & $0.77$ & $0.77$ & ~$0.76$ & ~~$-$~~ & ~~$\times$~~ 
\\
$Z_{I,II{\rm min}}$  & $3.03$ & $3.05$ & $3.07$ & ~~$3.14$ & $3.23$ & $3.43$ & ~~$3.79$~ & $11.24$ & $\times$ 
\\
\hline\hline
  &     &   &   &  & ${\rm tan}\beta=30$  &    & &  & 
  \\
\hline 
$\lambda$  & {\bf $\framebox[1.3\width]{0.0}$} & {\bf $\framebox[1.3\width]{0.1}$} & {\bf $\framebox[1.3\width]{0.2}$} & ~{\bf $\framebox[1.3\width]{0.3}$} & {\bf $\framebox[1.3\width]{0.4}$} & {\bf $\framebox[1.3\width]{0.5}$} & {\bf $\framebox[1.3\width]{0.6}$} & $0.7$ & $0.75$  
\\
\hline 
$y_{N{\rm max}}$  & ~~$0.86$~~ & ~~$0.86$~~ & ~~$0.85$~~ & ~~$0.84$ & $0.83$ & $0.81$ & ~~$0.78$~~ & ~~$0.37$~~ & ~~$\times$~~ 
\\
$y_{N{\rm min}}$  & ~~$0.77$~~ & ~~$0.77$~~ & ~~$0.77$~~ & ~~$0.77$ & $0.77$ & $0.77$ & ~~$0.76$~~ & ~~$-$~~ & ~~$\times$~~ 
\\
$Z_{I,II{\rm min}}$  & $2.95$ & $2.96$ & $3.01$ & ~~$3.11$ & $3.25$ & $3.51$ & $4.07$ & $80.01$ & $\times$ 
\\
\hline\hline
  &     &   &   &  & ${\rm tan}\beta=40$  &    & &  & 
  \\
\hline 
$\lambda$  & {\bf $\framebox[1.3\width]{0.0}$} & {\bf $\framebox[1.3\width]{0.1}$} & {\bf $\framebox[1.3\width]{0.2}$} & ~{\bf $\framebox[1.3\width]{0.3}$} & {\bf $\framebox[1.3\width]{0.4}$} & {\bf $\framebox[1.3\width]{0.5}$} & $0.6$ & $0.7$ & $0.75$  
\\
\hline 
$y_{N{\rm max}}$  & ~~$0.85$~~ & ~~$0.85$~~ & ~~~$0.85$~~ & ~~$0.84$ & $0.82$ & $0.80$ & ~~~$0.66$~~ & ~~$\times$~~ & ~~$\times$~~ 
\\
$y_{N{\rm min}}$  & ~~$0.77$~~ & ~~$0.77$~~ & ~~~$0.77$~~ & ~~$0.77$ & $0.77$ & $0.76$ & ~~~$-$~~ & ~~$\times$~~ & ~~$\times$~~ 
\\
$Z_{I,II{\rm min}}$  & $2.98$ & $3.01$ & ~$3.06$ & ~~$3.18$ & $3.37$ & $3.76$ & ~$8.05$ & $\times$ & $\times$ 
\\
\hline\hline
  &     &   &   &  & ${\rm tan}\beta=50$  &    & &  & 
  \\
\hline 
$\lambda$  & {\bf $\framebox[1.3\width]{0.0}$} & {\bf $\framebox[1.3\width]{0.1}$} & {\bf $\framebox[1.3\width]{0.2}$} & ~{\bf $\framebox[1.3\width]{0.3}$} & {\bf $\framebox[1.3\width]{0.4}$} & $0.5$ & $0.6$ & $0.7$ & $0.75$  
\\
\hline 
$y_{N{\rm max}}$  & ~~$0.84$~~ & ~~$0.84$~~ & ~~~$0.83$~~ & ~~$0.82$ & $0.80$ & $0.62$ & ~~$\times$~~ & ~~$\times$~~ & ~~$\times$~~ 
\\
$y_{N{\rm min}}$  & ~~$0.77$~~ & ~~$0.77$~~ & ~~~$0.77$~~ & ~~$0.77$ & $0.77$ & $-$ & ~~$\times$~~ & ~~$\times$~~ & ~~$\times$~~ 
\\
$Z_{I,II{\rm min}}$  & $3.12$ & $3.16$ & ~$3.23$ & ~~$3.42$ & $3.78$ & $10.40$ & $\times$ & $\times$ & $\times$ 
\end{tabular}
\end{center}\caption{Maximally allowed low energy values of $y_N$ ($=y_{N{\rm max}}$) and the corresponding minimal values of $Z_{I,II}$ for $\widetilde{m}_t=700~{\rm GeV}$, $q=\sqrt{5}$, and $b_{Z^\prime} =14q^2$. 
$y_{N{\rm min}}$s are $y_N$s needed for $Z_{I,II}=4.5$, yielding the 126 GeV Higgs mass. 
}\label{tab:LocalU(1)}
\end{table}
%
%

\section{Conclusion} \label{sec:conclusion}

The observed Higgs mass and the LP constraint seriously restrict the valid ranges of $\lambda$ and ${\rm tan}\beta$ in the NMSSM, 
leaving only the narrow bands, 
$0.6\lesssim\lambda\lesssim 0.7$ and $1<{\rm tan}\beta\lesssim 3$. 
Here the lower bound of $\lambda$ and the upper bound of ${\rm tan}\beta$ result from the 126 GeV Higgs mass.
In order to relieve such severe bounds, we extended the NMSSM with the vector-like superfields $\{H_d,H_u;N^c,N\}$, 
and studied their coupling with the MSSM Higgs doublet, $W=y_NN^ch_uH_d+\cdots$. 
We introduced also a U(1) gauge symmetry, 
under which only the extra vector-like superfields 
are charged, but all the ordinary NMSSM superfields remain neutral. 
With the help of such a U(1) gauge symmetry, 
the allowed value of $y_N$ at low energy can be lifted up to $0.85$, evading a LP below the GUT scale. 

The $y_N$ term and the holomorphic soft terms can remarkably raise the radiative Higgs mass particularly for large values of ${\rm tan}\beta$. 
Consequently, they invalidate the previous lower bound of $\lambda$ and the upper bound of ${\rm tan}\beta$, significantly enlarging the valid parameter space. 
In particular, the lower bound of $\lambda$ is completely removed for ${\rm tan}\beta\gtrsim 8$.
Thus, we have $0\lesssim\lambda \lesssim 0.4$-$0.6$ for $8\lesssim {\rm tan}\beta\lesssim 50$ as a consistent parameter space, 
while $0.4$-$0.5\lesssim\lambda \lesssim 0.6$-$0.7$ for $4\lesssim {\rm tan}\beta\lesssim 6$, 
and roughly $0.75\lesssim y_N \lesssim 0.85$, depending on ${\rm tan}\beta$. 
For $2\lesssim {\rm tan}\beta\lesssim 4$, 
the effects coming from the extra matter become weaker, 
and so we have just a limited parameter range, $0.5\lesssim\lambda\lesssim 0.6$. 
However, the original NMSSM parameter space should be contained in our case, and so relatively smaller $y_N$s in $0\lesssim y_N\lesssim 0.75$  
are also possible in small ${\rm tan}\beta$ cases.


\acknowledgments

\noindent 
I thank Jihn E. Kim for valuable discussion.  
This research is supported by Basic Science Research Program through the 
National Research Foundation of Korea (NRF) funded by the Ministry of Education, Grant No. 2013R1A1A2006904.


\end{document}